\documentclass[twocolumn,times,twocolappendix]{aastex631}

\usepackage{amsmath}

\newcommand{\Msun}{\ensuremath{M_\odot}}
\newcommand{\Rsun}{\ensuremath{R_\odot}}

\def\vsh{\ensuremath{v_{\rm sh}}}
\def\vsn{\ensuremath{v_{\rm SN}}}
\def\vcsm{\ensuremath{v_{\rm CSM}}}
\def\Rsh{\ensuremath{R_{\rm sh}}}
\def\Msh{\ensuremath{M_{\rm sh}}}

\def\Angstrom{\textup{\AA}}

\newcommand{\CfA}{\affiliation{Center for Astrophysics \textbar{} Harvard \& Smithsonian, 60 Garden Street, Cambridge, MA 02138-1516, USA}}
\newcommand{\IAIFI}{\affiliation{The NSF AI Institute for Artificial Intelligence and Fundamental Interactions, USA}}
\newcommand{\LCO}{\affiliation{Las Cumbres Observatory, 6740 Cortona Drive, Suite 102, Goleta, CA 93117-5575, USA}}
\newcommand{\UCSB}{\affiliation{Department of Physics, University of California, Santa Barbara, CA 93106-9530, USA}}

\newcommand{\STSci}{\affiliation{Space Telescope Science Institute, 3700 San Martin Drive, Baltimore, MD 21218, USA}}

\newcommand{\UA}{\affiliation{Steward Observatory, University of Arizona, 933 North Cherry Avenue, Tucson, AZ 85721-0065, USA}}
\newcommand{\Catalyst}{\altaffiliation{LSSTC Catalyst Fellow}}

\newcommand{\Columbia}{\affiliation{Department of Physics and Columbia Astrophysics Laboratory, Columbia University, 538 West 120th Street, New York, NY 10027, USA }}

\newcommand{\PennState}{\affiliation{Department of Astronomy \& Astrophysics, The Pennsylvania State University, 525 Davey Laboratory, University Park, PA 16802, USA}}
\newcommand{\PennStateICD}{\affiliation{Institute for Computational \& Data Sciences, The Pennsylvania State University, University Park, PA 16802, USA}}
\newcommand{\PennStateIGC}{\affiliation{Institute for Gravitation and the Cosmos, The Pennsylvania State University, University Park, PA 16802, USA}}
\newcommand{\MIT}{\affiliation{MIT-Kavli Institute for Astrophysics and Space Research, 77 Massachusetts Avenue, Cambridge, MA 02139, USA}}
\newcommand{\RESCEU}{\affiliation{Research Center for the Early Universe, Graduate School of Science, The University of Tokyo, 7-3-1 Hongo, Bunkyo-ku, Tokyo 113-0033, Japan}}

\shorttitle{Sharp Multi-Peak Type IIn SN 2021qqp with a Long Precursor}
\shortauthors{Hiramatsu et al.}

\begin{document}

\title{\bf \Large Multiple Peaks and a Long Precursor in the Type IIn Supernova 2021qqp: An Energetic Explosion in a Complex Circumstellar Environment}

\author[0000-0002-1125-9187]{Daichi Hiramatsu}
\CfA\IAIFI

\correspondingauthor{Daichi~Hiramatsu}
\email{daichi.hiramatsu@cfa.harvard.edu}

\author[0000-0002-9350-6793]{Tatsuya~Matsumoto}
\Columbia

\author[0000-0002-9392-9681]{Edo~Berger}
\CfA\IAIFI

\author[0000-0003-4175-4960]{Conor~Ransome}
\PennState

\author[0000-0002-5814-4061]{V.~Ashley~Villar}
\PennState\PennStateICD\PennStateIGC

\author[0000-0001-6395-6702]{Sebastian~Gomez}
\STSci

\author[0000-0001-7007-6295]{Yvette~Cendes}
\CfA

\author[0000-0002-8989-0542]{Kishalay~De}
\altaffiliation{NASA Einstein Fellow}
\MIT

\author[0000-0002-4924-444X]{K.~Azalee~Bostroem}
\Catalyst\UA

\author[0000-0003-4914-5625]{Joseph~Farah}
\LCO\UCSB

\author[0000-0003-4253-656X]{D.~Andrew~Howell}
\LCO\UCSB

\author[0000-0001-5807-7893]{Curtis~McCully}
\LCO\UCSB

\author[0000-0001-9570-0584]{Megan~Newsome}
\LCO\UCSB

\author[0000-0003-0209-9246]{Estefania~Padilla~Gonzalez}
\LCO\UCSB

\author[0000-0002-7472-1279]{Craig~Pellegrino}
\LCO\UCSB

\author[0000-0002-7043-6112]{Akihiro~Suzuki}
\RESCEU

\author[0000-0003-0794-5982]{Giacomo~Terreran}
\LCO\UCSB

\begin{abstract}
We present optical photometry and spectroscopy of the Type IIn supernova (SN) 2021qqp. Its unusual light curve is marked by a long precursor for $\approx300$\,days, a rapid increase in brightness for $\approx60$\,days, and then a sharp increase of $\approx1.6$\,mag in only a few days to a first peak of $M_r \approx -19.5$\,mag. The light curve then declines rapidly until it re-brightens to a second distinct peak of $M_r \approx -17.3$\,mag centered at $\approx335$\,days after the first peak. The spectra are dominated by Balmer lines with a complex morphology, including a narrow component with a width of $\approx 1300 \, {\rm km \, s}^{-1}$ (first peak) and $\approx 2500 \, {\rm km \, s}^{-1}$ (second peak) that we associate with the circumstellar medium (CSM) and a P Cygni component with an absorption velocity of $\approx 8500 \, {\rm km \, s}^{-1}$ (first peak) and $\approx 5600\, {\rm km \, s}^{-1}$ (second peak) that we associate with the SN--CSM interaction shell. Using the luminosity and velocity evolution, we construct a flexible analytical model, finding two significant mass-loss episodes with peak mass loss rates of $\approx 10$ and $\approx 5\, \Msun \, {\rm yr}^{-1}$ about $0.8$ and $2$\,yr before explosion, respectively, with a total CSM mass of $\approx 2-4\,\Msun$. We show that the most recent mass-loss episode could explain the precursor for the year preceding the explosion. The SN ejecta mass is constrained to be $\approx 5-30\,\Msun$ for an explosion energy of $\approx (3-10)\times10^{51}\,{\rm erg}$. We discuss eruptive massive stars (luminous blue variable, pulsational pair instability) and an extreme stellar merger with a compact object as possible progenitor channels.
\end{abstract}

\keywords{
\href{https://vocabs.ardc.edu.au/repository/api/lda/aas/the-unified-astronomy-thesaurus/current/resource.html?uri=http://astrothesaurus.org/uat/1668}{Supernovae (1668)}; 
\href{https://vocabs.ardc.edu.au/repository/api/lda/aas/the-unified-astronomy-thesaurus/current/resource.html?uri=http://astrothesaurus.org/uat/304}{Core-collapse supernovae (304)}; 
\href{https://vocabs.ardc.edu.au/repository/api/lda/aas/the-unified-astronomy-thesaurus/current/resource.html?uri=http://astrothesaurus.org/uat/1731}{Type II supernovae (1731)}; 
\href{https://vocabs.ardc.edu.au/repository/api/lda/aas/the-unified-astronomy-thesaurus/current/resource.html?uri=http://astrothesaurus.org/uat/732}{Massive stars (732)};
\href{https://vocabs.ardc.edu.au/repository/api/lda/aas/the-unified-astronomy-thesaurus/current/resource.html?uri=http://astrothesaurus.org/uat/1613}{Stellar mass loss (1613)};
\href{https://vocabs.ardc.edu.au/repository/api/lda/aas/the-unified-astronomy-thesaurus/current/resource.html?uri=http://astrothesaurus.org/uat/241}{Circumstellar matter (241)}
}

\section{Introduction} 
\label{sec:intro}

Among hydrogen-rich (H-rich) Type II supernovae (SNe~II), Type IIn SNe (SNe~IIn) are classified based on the presence of ``narrow'' Balmer-series emission lines in their spectra \citep{Schlegel1990MNRAS.244..269S,Filippenko1997ARA&A..35..309F}. The main power source of SNe IIn is thought to be the shock interaction between the SN ejecta and circumstellar material (CSM; see \citealt{Smith2017hsn..book..403S} for a review), resulting in the most heterogeneous SN class in terms of their observed properties (e.g., light curves and spectra; see \citealt{Taddia2013A&A...555A..10T,Nyholm2020A&A...637A..73N} for sample studies).  The nature of the underlying SNe and their progenitors remains elusive, as their observational signatures are mostly hidden below the photosphere formed at the CSM interaction layer.

The direct progenitor identification of SN~IIn 2005gl in a pre-explosion image points a massive ($>50\,M_\odot$) luminous blue variable (LBV) as a potential progenitor for that event \citep{Gal-Yam2007ApJ...656..372G,Gal-Yam2009Natur.458..865G}. Given the observed heterogeneous properties of SNe~IIn, however, various other progenitor channels are also possible, for example, super-asymptotic giant branch stars ($\sim8$--$10\,M_\odot$; e.g., \citealt{Kankare2012MNRAS.424..855K,Mauerhan2013MNRAS.431.2599M,Smith2013MNRAS.434..102S,Moriya2014A&A...569A..57M,Hiramatsu2021NatAs...5..903H}), extreme red supergiants ($\sim17$--$25\,M_\odot$; e.g., \citealt{Fullerton2006ApJ...637.1025F,Smith2009AJ....137.3558S,Moriya2011MNRAS.415..199M,Mauerhan2012MNRAS.424.2659M}), interacting massive binaries ($>20\,M_\odot$; e.g., \citealt{Chevalier2012ApJ...752L...2C,Soker2013ApJ...764L...6S,Kashi2013MNRAS.436.2484K,Schroder2020ApJ...892...13S,Metzger2022ApJ...932...84M}), and even pulsational pair-instability SNe (PPISNe; $\sim110-140\,\Msun$; e.g.,  \citealt{Woosley2007Natur.450..390W,Blinnikov2010PAN....73..604B,Moriya2013MNRAS.428.1020M,Woosley2017ApJ...836..244W}).

The observed heterogeneity reflects the diversity in pre-explosion mass loss responsible for the CSM formation. Therefore, SNe~IIn with precursor events provide a unique opportunity to directly connect mass-loss activity to the resultant SN properties (e.g., \citealt{Ofek2013Natur.494...65O}; see \citealt{Ofek2014ApJ...789..104O,Bilinski2015MNRAS.450..246B,Strotjohann2021ApJ...907...99S} for sample studies).  One of the most well-observed examples is SN~2009ip, which was discovered during its LBV-like giant eruption (i.e., SN impostor) phase in 2009 \citep{Smith2010AJ....139.1451S,Foley2011ApJ...732...32F}, followed by a more luminous SN~IIn-like event in 2012 \citep{Prieto2013ApJ...763L..27P, Mauerhan2013MNRAS.430.1801M,Pastorello2013ApJ...767....1P,Fraser2013MNRAS.433.1312F,Margutti2014ApJ...780...21M,Levesque2014AJ....147...23L,Smith2014MNRAS.438.1191S,Graham2014ApJ...787..163G,Mauerhan2014MNRAS.442.1166M,Martin2015AJ....149....9M,Fraser2015MNRAS.453.3886F,Graham2017MNRAS.469.1559G,Reilly2017MNRAS.470.1491R,Smith2022MNRAS.515...71S}. Multi-peak light curves seen in some precursor-associated SNe~IIn (e.g., iPTF13z; \citealt{Nyholm2017A&A...605A...6N}, as well as SN~2009ip) also suggest complex CSM structures from eruptive, rather than steady, mass loss.

Here, we report detailed optical photometry and spectroscopy of SN~IIn 2021qqp, which exhibits clear precursor activity directly up to the SN explosion and multiple peaks indicative of distinct eruptive mass-loss episodes.  We further construct an analytical model to directly extract the CSM and SN properties from the combined light-curve and spectral properties. The paper is structured as follows. In \S\ref{sec:disc} and \ref{sec:obs}, we summarize the discovery, classification, archival and follow-up observations, and data reduction; in \S\ref{sec:ana}, we analyze the host galaxy and SN light curves and spectra;
we present an analytical model to extract the CSM and SN properties in \S\ref{sec:mod};
and we discuss possible progenitor channels in \S\ref{sec:sum} and conclude with a future outlook in \S\ref{sec:conc}.

\section{Discovery and Classification} 
\label{sec:disc}

The Zwicky Transient Facility (ZTF; \citealt{Bellm2019PASP..131a8002B, Graham2019PASP..131g8001G}) discovered SN~2021qqp (ZTF21abgjldn) on 2021 May 23.47 (UT dates are used throughout; ${\rm MJD} = 59357.47$) at an $r$-band magnitude of 20.86 at $\text{R.A.}=22^{\text{h}}32^{\text{m}}40^{\text{s}}.419$ and $\text{decl.}=+25^{\circ}34'34".79$ \citep{De2021TNSTR2150....1D}. 
Subsequent discoveries have been reported by the Asteroid Terrestrial-impact Last Alert System (ATLAS; \citealt{Tonry2018PASP..130f4505T, Smith2020PASP..132h5002S}) and Pan-STARRS1 (PS1; \citealt{Chambers2016arXiv161205560C}), with the PS1 detection being the earliest, on 2020 December 24.22 (${\rm MJD} = 59207.22$; $150$ days before discovery), with an $i$-band magnitude of 21.39 at $\text{R.A.}=22^{\text{h}}32^{\text{m}}40^{\text{s}}.416$ and $\text{decl.}=+25^{\circ}34'34".75$. Given the smallest pixel scale, we adopt the PS1 coordinates in this work.

\cite{Chu2021TNSCR2605....1C} obtained an optical spectrum of SN~2021qqp on 2021 July 6.51 (${\rm MJD} = 59401.51$; $44$ days after discovery) with the Low Resolution Imaging Spectrometer (LRIS; \citealt{Oke1995PASP..107..375O,McCarthy1998SPIE.3355...81M,Rockosi2010SPIE.7735E..0RR}) mounted on the 10\,m Keck I Telescope (Hawaii, USA), classifying it as an SN~IIn at $z=0.041$ that is consistent with the host galaxy redshift of $z=0.041475\pm0.000087$  (ALFALFA 4-043; \citealt{Martin2009ApJS..183..214M,Haynes2011AJ....142..170H,Haynes2018ApJ...861...49H}).\footnote{Via the NASA/IPAC Extragalactic Database: \url{http://ned.ipac.caltech.edu/}}
A subsequent spectrum was obtained on 2021 July 29.24 (${\rm MJD} = 59424.24$; $67$ days after discovery) by the Public European Southern Observatory (ESO) Spectroscopic Survey for Transient Objects \citep{Smartt2015A&A...579A..40S} using the ESO  Faint Object Spectrograph and Camera (EFOSC2; \citealt{Buzzoni1984Msngr..38....9B}) mounted on the $3.58$\,m New Technology Telescope (NTT; La Silla, Chile), confirming the SN IIn classification and redshift \citep{Munoz2021TNSCR2620....1M,Munoz2021TNSAN.204....1M}.

In this work, we assume a standard $\Lambda$CDM cosmology with $H_0=71.0$\, km\,s$^{-1}$\,Mpc$^{-1}$, $\Omega_{\Lambda}=0.7$, and $\Omega_m=0.3$ and convert the redshift to a luminosity distance of $d_L=181$\,Mpc (distance modulus, $\mu=36.28$\,mag).

\section{Observation and Data Analysis} 
\label{sec:obs}

\subsection{Optical and Infrared Photometry}
\label{sec:phot}

Through the Global Supernova Project \citep{Howell2017AAS...23031803H}, we obtained Las Cumbres Observatory (LCO; \citealt{Brown2013PASP..125.1031B}) $BgVri$-band imaging data with the Sinistro cameras on the network of 1\,m telescopes at the Cerro Tololo Inter-American Observatory (District IV, Chile), McDonald Observatory (Texas, USA), South African Astronomical Observatory (Sutherland, South Africa), and Teide Observatory (Canary Islands, Spain), as well as $griz$-band imaging data with the Multicolor Simultaneous Camera for studying Atmospheres of Transiting exoplanets 3 (MuSCAT3; \citealt{Narita2020SPIE11447E..5KN}) on the 2\,m Faulkes Telescope North (Hawaii, USA) from 2021 November 17 to 2022 November 25 (${\rm MJD}=59535-59908$). 
LCO photometry was performed with point-spread function (PSF) fitting using \texttt{lcogtsnpipe}\footnote{\url{https://github.com/LCOGT/lcogtsnpipe}} \citep{Valenti2016MNRAS.459.3939V}, a PyRAF-based photometric reduction pipeline. $BV$- and $griz$-band data were calibrated to Vega and AB magnitudes, respectively, using the 9th Data Release of the AAVSO Photometric All Sky Survey \citep{Henden2016yCat.2336....0H} and the 13th Data Release of the Sloan Digital Sky Survey (SDSS; \citealt{Albareti2017ApJS..233...25A}).

To explore possible pre-explosion variability (as indicated by the first PS1 detection being $\approx 5$ months before the ZTF discovery; \S\ref{sec:disc}) and to obtain additional post-explosion photometry of SN~2021qqp, we process and examine ZTF, ATLAS, PS1, Palomar Transient Facility (PTF; \citealt{Law2009PASP..121.1395L}), and Wide-field Infrared Survey Explorer (WISE; \citealt{Wright2010AJ....140.1868W,Mainzer2014ApJ...792...30M}) survey data. ZTF and ATLAS photometry was directly retrieved from, respectively, the ZTF forced-photometry service\footnote{\url{https://ztfweb.ipac.caltech.edu/cgi-bin/requestForcedPhotometry.cgi}} \citep{Masci2019PASP..131a8003M} in the $g$, $r$, and $i$ bands (date range: 2018 April 26 to 2023 January 12; ${\rm MJD}=58234-59956$), and the ATLAS forced photometry server\footnote{\url{https://fallingstar-data.com/forcedphot/}} \citep{Shingles2021TNSAN...7....1S} in the $c$ and $o$ bands (date range: 2015 August 6 to 2023 January 21; ${\rm MJD}=57240-59965$). 

We retrieved PS1 and PTF single-epoch and co-added reference images from the PS1 Image Cutout Service\footnote{\url{http://ps1images.stsci.edu/cgi-bin/ps1cutouts}} \citep{Flewelling2020ApJS..251....7F} in the $g$, $r$, and $i$ bands (date range: 2010 August 16 to 2014 October 12; ${\rm MJD}=55424-56942$) and PTF NASA/IPAC Infrared Science Archive (IRSA)\footnote{\url{https://irsa.ipac.caltech.edu/Missions/ptf.html}} in the $g$ and $R$ bands (date range: 2010 July 26 to 2014 October 30; ${\rm MJD}=55403-56944$), respectively. The data processing, combining, and scaling processes used are described by \citet{Magnier2020ApJS..251....3M} and \citet{Waters2020ApJS..251....4W} for PS1 and \citet{Ofek2012PASP..124...62O} for PTF. In both cases, during the stacking process, the single-epoch images have a scaling factor applied such that the stacked image zero-point magnitudes are 25 for PS1 and 27 for PTF. We used the scaling factor to scale the single-epoch images to the stacked image and subtract the co-added image from the single epochs to remove host galaxy light. Forced aperture photometry was performed using standard \texttt{photutils} \citep[v1.7.0;][]{photutils} routines with aperture sizes representative of the typical FWHM of each survey (2" for PTF and 2\farcs5 for PS1).

The SN field has also been observed by the ongoing NEOWISE all-sky mid-IR survey in the $W1$ ($3.4$\,$\mu$m) and $W2$ ($4.5$\,$\mu$m) bands \citep{Wright2010AJ....140.1868W, Mainzer2014ApJ...792...30M}. We retrieved time-resolved co-added images of the field created as part of the unWISE project \citep{Lang2014AJ....147..108L, Meisner2018AJ....156...69M}. To remove contamination from the host galaxy, we used a custom code \citep{De2020PASP..132b5001D} based on the ZOGY algorithm \citep{Zackay2016ApJ...830...27Z} to perform image subtraction on the NEOWISE images using the full-depth coadds of the WISE and NEOWISE mission (obtained during 2010--2014) as reference images. Photometric measurements were obtained by performing forced PSF photometry at the transient position on the subtracted WISE images until the epoch of the unWISE data release (data acquired until 2021 December).

The detection significance ($\sigma$) of the ZTF, ATLAS, PS1, PTF, and WISE forced photometry was determined from the ratio of measured flux ($f$) to its error ($f_{\rm err}$). For the measurements above and below $3\sigma$, we report their detections ($-2.5\,{\rm log}_{10}(f)+{\rm ZP}$) and $3\sigma$ upper limits ($-2.5\,{\rm log}_{10}(3\times f_{\rm err})+{\rm ZP}$), respectively, where ``ZP'' is the zero-point in the AB magnitude system.

\begin{figure*}
    \centering
    \includegraphics[width=1\textwidth]{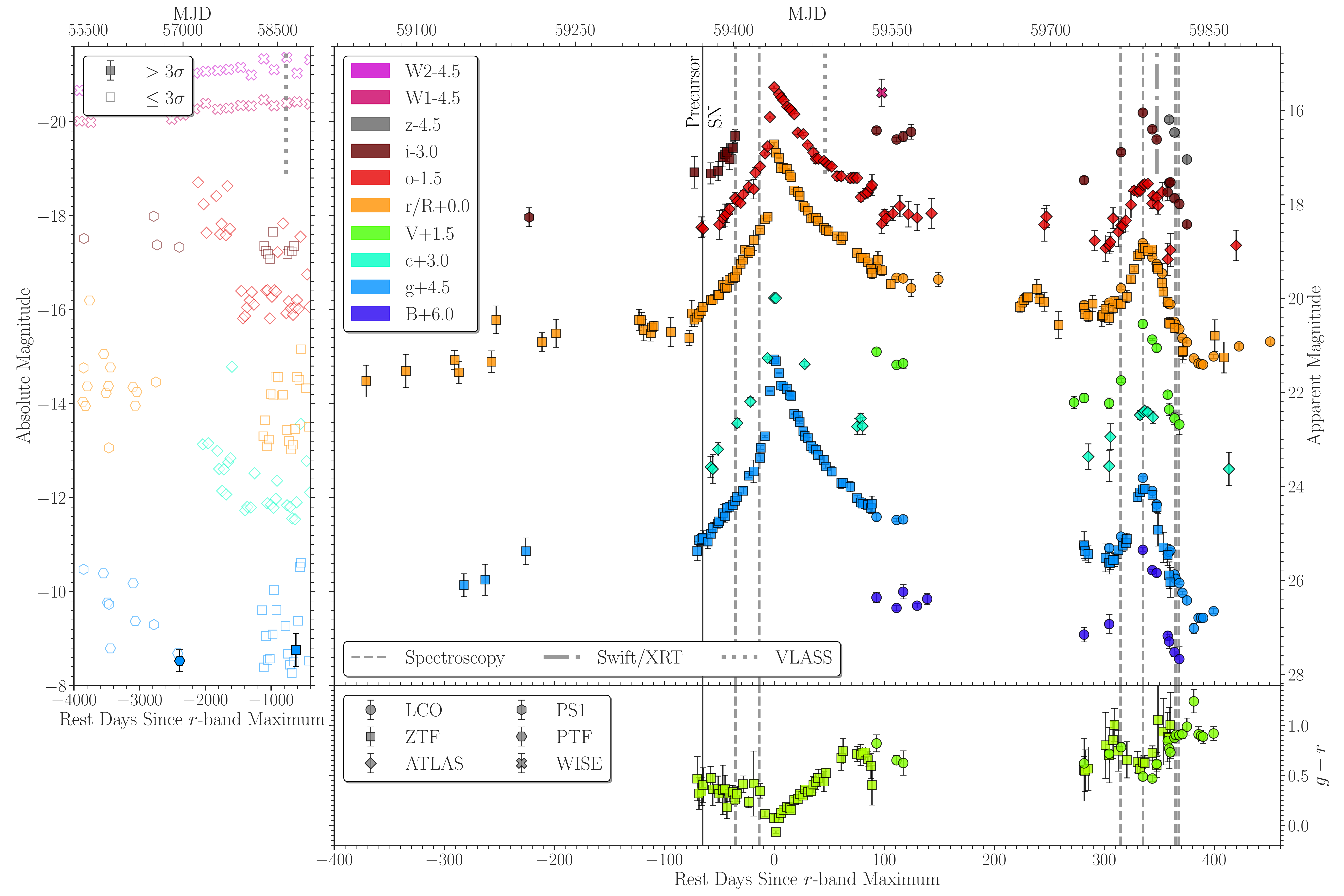}
\caption{Multi-band light curves ({\it Top}) and $g-r$ color evolution ({\it Bottom}) of SN~2021qqp. Filled and open symbols in the left panel are used for detections and $3\sigma$ upper limits (binned every 30\,days), respectively. Error bars denote $1\sigma$ uncertainties and are sometimes smaller than the marker size. The data gaps are due to Sun observing constraints. The gray vertical dashed lines mark the times of spectroscopic observations (Figure~\ref{fig:spec}); dotted and dashed-dotted lines mark radio and X-ray observations (\S~\ref{sec:xray}), respectively.  The light curve transitions from a gradual to a sharper rise at $\approx -65$\,days, corresponding to the precursor-to-SN transition (the black vertical line).   The $g-r$ color co-evolves with the light curve, with bluer colors as the light curve peaks. A few faint $\sim3\sigma$ detections are apparent up to $6.5$\,yr prior to the SN peak. (The data used to create this figure are available.)
}
    \label{fig:lc}
\end{figure*}

\begin{figure*}
    \centering
    \includegraphics[width=1\textwidth]{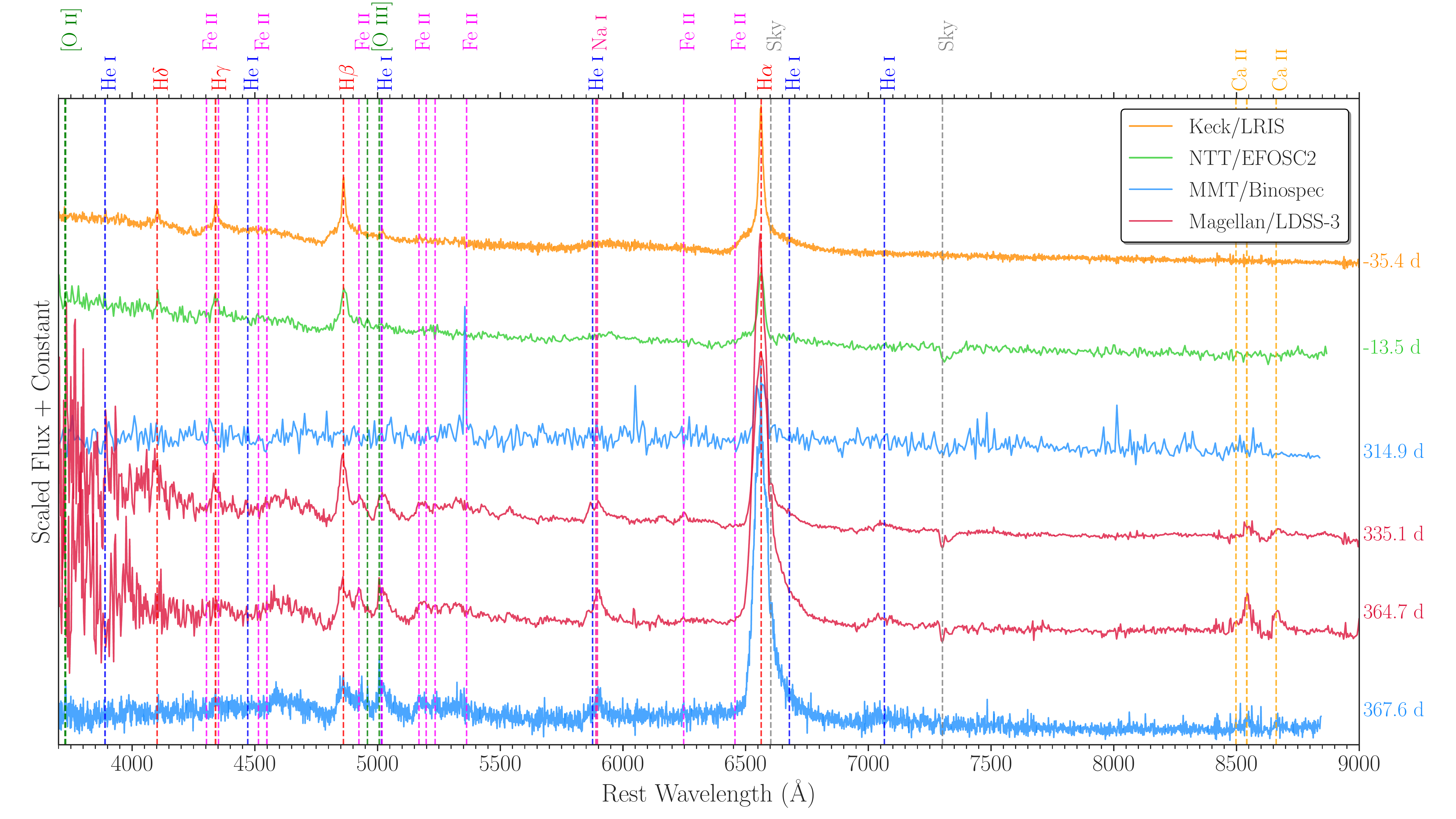}
\caption{Spectral time series of SN~2021qqp with the phases denoted on the right. The Balmer lines and blue continuum are seen in the first two spectra, while the weaker He~{\sc i}, Na~{\sc i}, Ca~{\sc i}, and Fe~{\sc ii} lines are also seen in the last four spectra as the continuum drops. The host galaxy [O~{\sc ii}] and [O~{\sc iii}] emission lines are detected in the final two spectra after the SN flux faded sufficiently (see also Figure~\ref{fig:HaHb}).
(The data used to create this figure are available.)
}
    \label{fig:spec}
\end{figure*}

\subsection{Optical Spectroscopy}
\label{sec:spec}

Through our FLEET program \citep{Gomez2020ApJ...904...74G,Gomez2023ApJ...949..114G}, we obtained optical spectra on 2022 July 6 and August 30 (${\rm MJD}=59766$ and $59821$) with Binospec \citep{Fabricant2019PASP..131g5004F} mounted on the 6.5 m MMT Observatory (Arizona, USA) and on 2022 July 27 and August 27 (${\rm MJD}=59787$ and $59818$) with the Low Dispersion Survey Spectrograph 3 (LDSS-3; \citealt{Stevenson2016ApJ...817..141S}) mounted on the 6.5\,m Magellan Clay Telescope (Cerro Manqui, Chile). The combinations of the 270 grating (Binospec) and VPH-All grism (LDSS-3) with a $1"$-long slit were used for dispersion, resulting in wavelength coverage of $3820-9210\, \Angstrom$ ($R\approx 1500$) and $3700-10060\,\Angstrom$ ($R\approx 700$), respectively. One-dimensional spectra were extracted, reduced, and calibrated following standard procedures using PyRAF and flux calibrated to a standard taken during the same week as the target spectra. Additionally, we retrieved the public Keck/LRIS and NTT/EFOSC2 classification spectra (\S\ref{sec:disc}) via the Transient Name Server (TNS)\footnote{\url{https://www.wis-tns.org/}} and include them in the subsequent analysis. Additional flux calibration was applied to all the spectra using coeval photometry.

All photometry and spectroscopy of SN~2021qqp are presented in Figures~\ref{fig:lc} and \ref{fig:spec}, respectively. No Na~{\sc i}~D absorption is seen at the host redshift (Figure~\ref{fig:spec}), indicating low host extinction at the SN position (Figure~\ref{fig:hostcircs}). Thus, we correct all photometry and spectroscopy only for the Milky Way (MW) extinction of $A_V=0.176$\,mag \citep{Schlafly2011ApJ...737..103S},\footnote{Via the NASA/IPAC IRSA: \url{https://irsa.ipac.caltech.edu/applications/DUST/}} assuming the \cite{Fitzpatrick1999PASP..111...63F} reddening law with $R_V=3.1$ and extended to the WISE bands with the relative optical-to-infrared extinction values from \cite{Wang2019ApJ...877..116W}. As SN~2021qqp is best sampled in the ZTF $r$ band, we use its epoch at maximum light (${\rm MJD}_{r,{\rm max}}=59438.33$) as the zero-point reference for all phases unless otherwise specified.

\subsection{X-Ray and Radio}
\label{sec:xray}

We obtained \textit{Neil G.~Gehrels Swift} X-Ray Telscope (XRT) observations on 2022 August 9 (${\rm MJD}=59800$; phase of $+348$\,days) with a total on-source exposure time of 3185 s.\footnote{UVOT observations were also obtained contemporaneously. As the SN signal is not detected in the UVOT images, we use them for the host-galaxy analysis in \S~\ref{sec:host}.} A $3\sigma$ upper limit of $4.4\times10^{-3}$\,counts\,s$^{-1}$ ($0.3-10$\,keV) was estimated using the \textit{Swift}-XRT web tool\footnote{\url{https://www.swift.ac.uk/user_objects/index.php}} \citep{Evans2007A&A...469..379E,Evans2009MNRAS.397.1177E}. With a MW H~{\sc i} column density of $4.6\times10^{20}$\,cm$^{-2}$ \citep{HI4PI2016A&A...594A.116H}\footnote{Via the NASA HEASARC N$_{\rm H}$ Tool: \url{https://heasarc.gsfc.nasa.gov/cgi-bin/Tools/w3nh/w3nh.pl}} and assuming a power-law spectrum with a photon index of 2, the count rate is converted\footnote{Via the NASA HEASARC WebPIMMS: \url{https://heasarc.gsfc.nasa.gov/cgi-bin/Tools/w3pimms/w3pimms.pl}} to an unabsorbed flux limit of $F_{\rm X}\lesssim 1.8\times10^{-13}$\,erg\,s$^{-1}$\,cm$^{-2}$, corresponding to $L_{\rm X}\lesssim6.8\times10^{41}$\,erg\,s$^{-1}$.  

We further obtained images at the location of the SN from the Very Large Array Sky Survey \citep[VLASS; ][]{Lacy2020PASP..132c5001L} and measured the flux density with the {\tt imtool fitsrc} command within {\tt pwkit} \citep{Williams2017ApJ...834..117W}. Two VLASS images exist, where the first was taken on 2019 May 26 (${\rm MJD}=58629$; phase of $-777$\,days) and the second on 2021 September 29 (${\rm MJD}=59486$; phase of $+46$\,days). In both cases, we place a $3\sigma$ upper limit of $\lesssim 0.3$\,mJy ($2-4$\,GHz), corresponding to $L_{\rm radio}\lesssim 3.4\times10^{37}$\,erg\,s$^{-1}$.  

The luminosity and temporal ranges probed by the X-ray and radio observations are not particularly constraining as compared to previous SN~IIn detections: $L_{\rm X}\sim10^{41}$ and $L_{\rm radio}\sim10^{37}$\,erg\,s$^{-1}$ at $\sim 1000$ days after explosion (e.g., \citealt{Chandra2018SSRv..214...27C}).

\section{Analysis} 
\label{sec:ana}

\subsection{Host Galaxy} 
\label{sec:host}

\begin{figure}
 \centering    
 \includegraphics[width=0.48\textwidth]{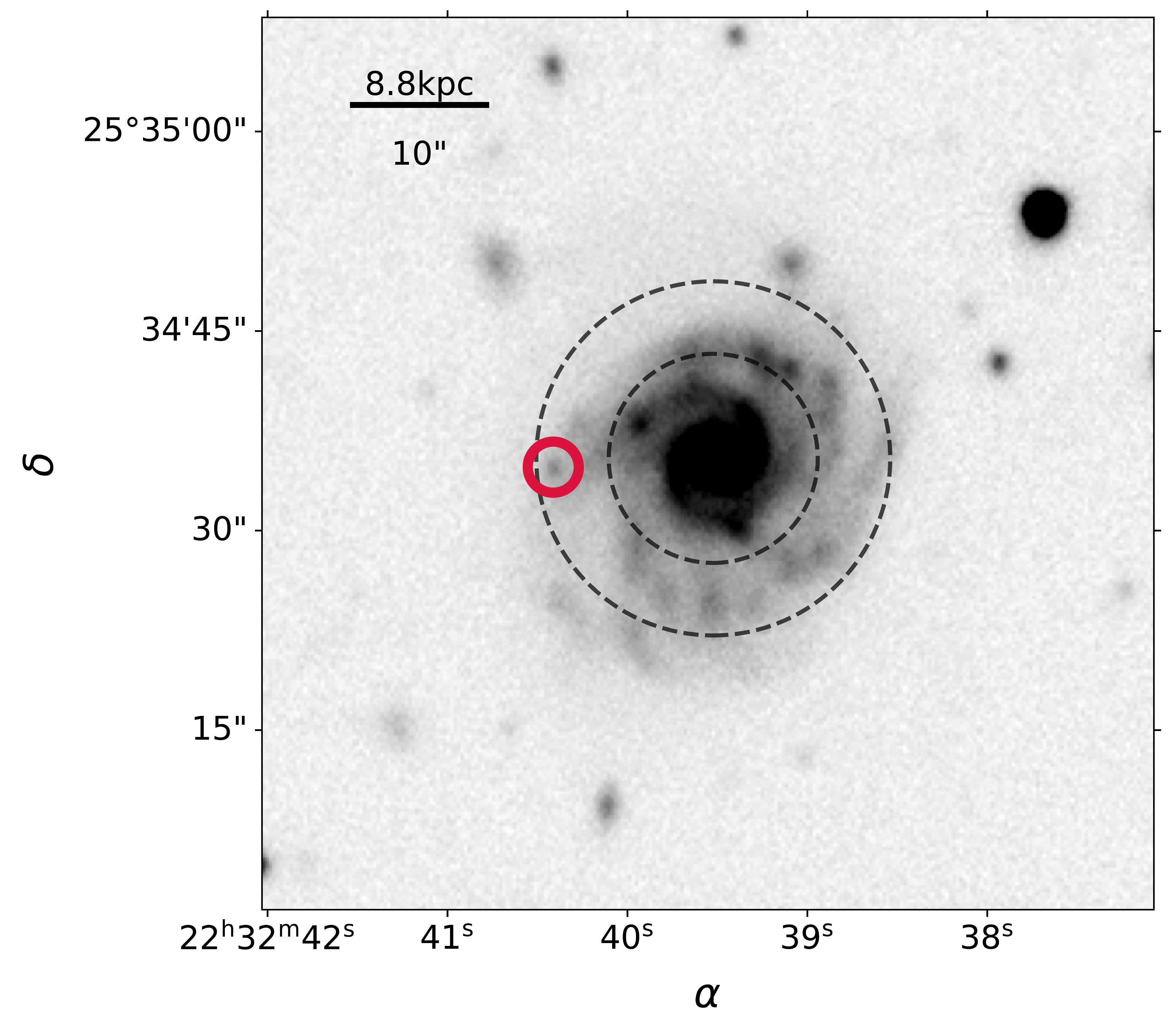}
\caption{The location of SN~2021qqp (red circle) relative to its host galaxy, indicating an association with a spiral arm. The dashed circles indicate $r_{50}=7\farcs9=6.9$\,kpc and $r_{80}=13\farcs3=11.7$\,kpc ($50\%$ and $80\%$ light radii), respectively. The image is in the $g$ band from the DESI Legacy Imaging Surveys \citep{DESILIS} Data Release 9.
}
    \label{fig:hostcircs}
\end{figure}

\begin{figure*}
 \centering    
 \includegraphics[width=0.84\textwidth]{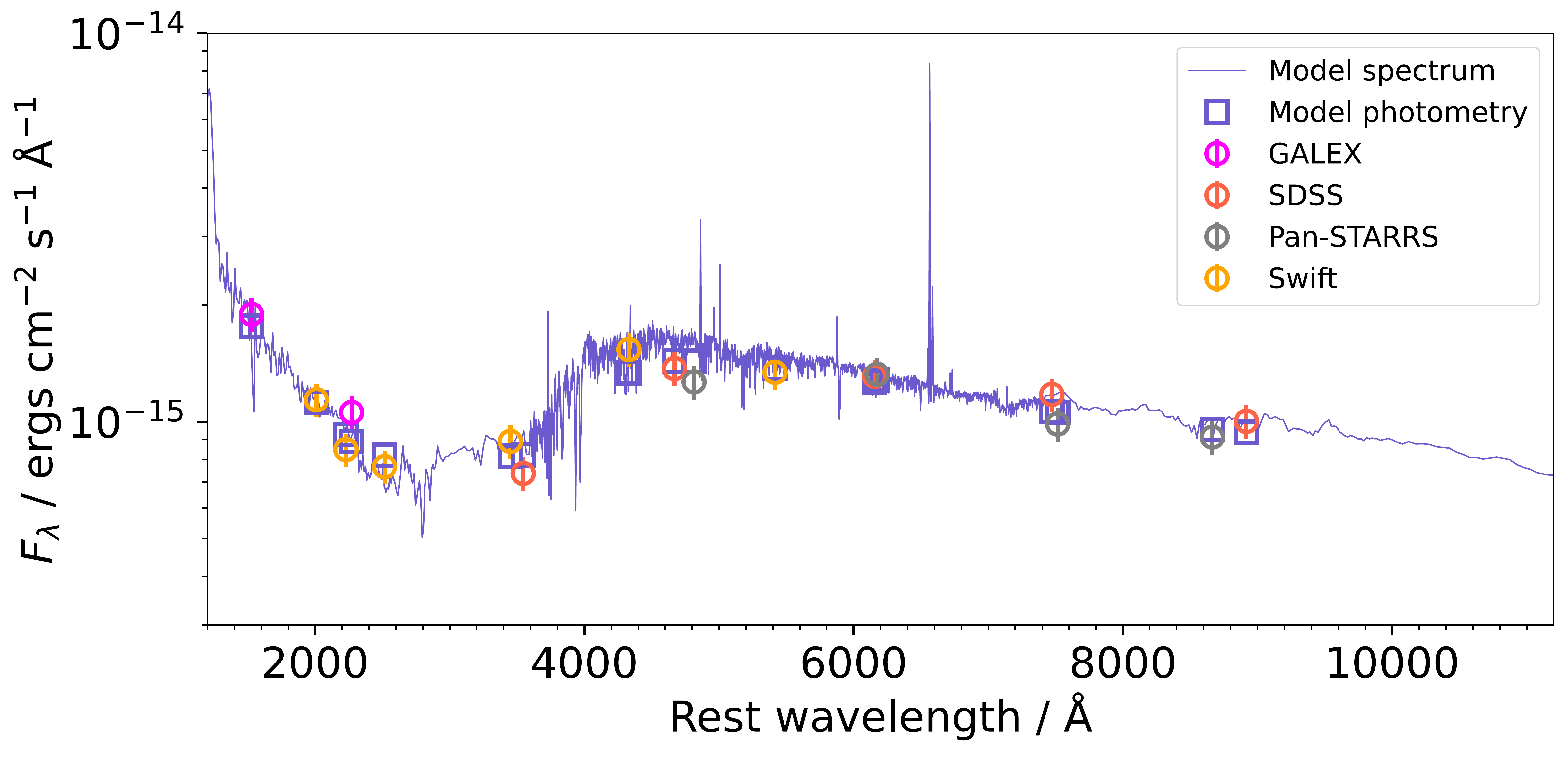}
\caption{Observed UV (GALEX+\textit{Swift}) and optical (\textit{Swift}+SDSS+PS1) fluxes for the host galaxy along with  \texttt{Prospector} model fits (blue line and squares). The inferred parameters from the model are listed in \S\ref{sec:host}. 
}
    \label{fig:hostsed}
\end{figure*}

The host galaxy of SN\,2021qqp is a face-on spiral galaxy,\footnote{\url{https://ned.ipac.caltech.edu/byname?objname=ALFALFA+4-043&hconst=67.8&omegam=0.308&omegav=0.692&wmap=4&corr_z=1}} as shown in Figure~\ref{fig:hostcircs}. SN\,2021qqp is offset from the center of its host by $\approx12\farcs2$, or $10.7$\,kpc with the assumed standard $\Lambda$CDM cosmology. As can be seen in Figure~\ref{fig:hostcircs}, SN\,2021qqp coincides with a spiral arm and is located just within the $r_{80}\approx 13\farcs3$ light radius. The location of SN\,2021qqp is not particularly unusual for SNe\,IIn \citep[e.g.][]{Galbany2014A&A...572A..38G,Galbany2016A&A...591A..48G,Galbany2018ApJ...855..107G,Schulze2021ApJS..255...29S,Ransome2022MNRAS.513.3564R}.
We note that there exists a cataloged PSF-like (i.e., stellar-like) object spatially coincident with the SN location (within $\approx0\farcs03$, smaller than a typical PSF width of $1\farcs2$, or $1.1$\,kpc) in the DESI Legacy Imaging Surveys (from 2014 February to 2019 March; ${\rm MJD}\approx56689-58573$; \citealt{DESILIS}) Data Release 9\footnote{\url{https://datalab.noirlab.edu/query.php?name=ls_dr9.tractor}} with the following AB magnitudes: $m_g=22.8$ ($M_g=-13.4$), $m_r=22.5$ ($M_r=-14.0$), and $m_z=23.1$ ($M_z=-13.3$). Given the luminosity, the cataloged object is unlikely to be a single quiescent star, but we cannot distinguish between a precursor activity (Figure~\ref{fig:lc}) and an unresolved star-forming region.

To estimate global host parameters such as stellar mass ($M_\star$), metallicity ($Z$), age ($t_{\rm age}$), and star formation rate (SFR), we use \texttt{Prospector}, a stellar population Bayesian inference package \citep{Johnson2021ApJS..254...22J} that has been extensively used to fit the spectral energy distributions (SEDs) of field galaxies and transient host galaxies (e.g., \citealt{Leja2017ApJ...837..170L, Blanchard2017ApJ...848L..22B, Nicholl2017ApJ...845L...8N,Schulze2021ApJS..255...29S}). \texttt{Prospector} fits photometry and/or spectra and creates an SED model. We fit photometry from the Galaxy Evolution Explore (GALEX; \citealt{Martin2005ApJ...619L...1M}) in the far-UV ($155$ nm) and near-UV ($230$ nm) bands; \textit{Swift} UV/Optical Telescope (UVOT)\footnote{The host-galaxy photometry was extracted with an aperture size of $r_{80}$.} in the $UVW2,UVM2,UVW1,U,B$, and $V$ bands; SDSS in the $u,g,r,i$, and $z$ bands; and PS1 in the $g,r,i$, and $z$ bands using appropriate prior distributions \citep[$M_\star$, $Z$, and $\tau$, a characteristic $e$-folding timescale of the delayed-$\tau$ star-formation history, ${\rm SFH} \propto t \times e^{-t/\tau}$; see][and references therein for details]{Carnall2019ApJ...873...44C} and nested sampling using \texttt{dynesty} \citep[v2.1.1;][]{Speagle2020MNRAS.493.3132S}. 

The fitting results are shown in Figure~\ref{fig:hostsed}, from which we find $M_\star = 1.74^{+0.64}_{-0.63}\times10^{10}\,M_\odot$, ${\rm log}_{10}(Z/Z_\odot)=-0.29^{+0.28}_{-0.34}$, $t_{\rm age}=8.98^{+3.24}_{-3.33}$\,Gyr, $\tau=1.86^{+0.83}_{-0.76}$\,Gyr, and a current SFR (=SFH($t_{\rm age}$)) of $0.59\,\pm\,0.22\,M_\odot\,{\rm yr}^{-1}$. Moreover, using the GALEX UV photometry, we estimate an SFR of $0.62\,\pm\,0.05\,M_\odot\,{\rm yr}^{-1}$ with the \citet{Salim2007ApJS..173..267S} calibration, which is consistent with the value from the SED fitting. The inferred $M_\star$ and current SFR are a few times smaller than those of the MW (e.g., \citealt{Licquia2015ApJ...806...96L}), but these global host properties are typical among SN~IIn hosts (e.g., \citealt{Schulze2021ApJS..255...29S}).

We also estimate a local SFR and metallicity from the host [O~{\sc ii}] and [O~{\sc iii}] emission lines detected in our final SN spectrum, when the SN flux faded sufficiently (Figures~\ref{fig:spec} and \ref{fig:HaHb}). We fit a double Gaussian profile to [O~{\sc ii}]\,$\lambda3727$ and [O~{\sc ii}]\,$\lambda3729$ and a single Gaussian profile each to [O~{\sc iii}]\,$\lambda4959$ and [O~{\sc iii}]\,$\lambda5007$. These fits yield luminosities of $L_{\rm [O~{\sc II}]\lambda3727}=(1.8\pm0.4)\times10^{38}$, $L_{\rm [O~{\sc II}]\lambda3729}=(1.9\pm0.3)\times10^{38}$, $L_{\rm [O~{\sc III}]\lambda4959}=(1.5\pm0.3)\times10^{38}$, and $L_{\rm [O~{\sc III}]\lambda5007}=(3.6\pm0.4)\times10^{38}$\,erg\,s$^{-1}$. 
Using the \cite{Kennicutt1998ARA&A..36..189K} SFR calibration with [O~{\sc ii}]\,$\lambda3727$ and the \cite{Maiolino2008A&A...488..463M} metallicity calibration with [O~{\sc iii}]\,$\lambda5007$/[O~{\sc ii}]\,$\lambda3727$, we estimate ${\rm SFR}_{\rm loc}=(2.5\pm0.9)\times10^{-3}\,M_\odot\,{\rm yr}^{-1}$ and ${\rm log}_{10}(Z_{\rm loc}/Z_\odot)=-0.33\pm0.09$, respectively. The local SFR and metallicity are on the low end of the distributions for SN~IIn local environements ($\lesssim10\%$ and $\lesssim20\%$ for SFR and metallicity, respectively; \citealt{Galbany2014A&A...572A..38G,Galbany2016A&A...591A..48G,Galbany2018ApJ...855..107G}).

\subsection{Light-curve Evolution} 
\label{sec:lc}

As shown in Figure~\ref{fig:lc}, the multi-band light curve of SN~2021qqp shows a gradual rise ($\approx -4$\,mmag\,day$^{-1}$) from $-14.5$\,mag in the $r$ band at $-370$\,days (or possibly even $-2400$\,days, although other transient events in the unresolved star-forming region (\S\ref{sec:host}) cannot be ruled out without a clear rising trend) with possible bumps around $-240$ and $\lesssim-120$\,days (during the Sun constraint). This ``precursor'' is similar in brightness to the Great Eruption of Eta Carinae and some SN impostors (e.g., \citealt{Humphreys1994PASP..106.1025H,Davidson1997ARA&A..35....1D,VanDyk2012ASSL..384..249V,Smith2017RSPTA.37560268S}) and likely caused by a pre-explosion mass-loss event(s). The precursor absolute magnitude and duration of SN~2021qqp ($-15.8\,{\rm mag} \lesssim M_{g,r,i} \lesssim -14.5\,{\rm mag}$ and $\approx 300$\,days) are on the luminous and long-lasting ends of the SN~IIn precursor distributions, respectively \citep{Ofek2014ApJ...789..104O,Strotjohann2021ApJ...907...99S}. By comparing with the estimated rates of precursor luminosity and duration from \cite{Strotjohann2021ApJ...907...99S} using 18 SNe~IIn discovered by ZTF with observed precursors brighter than $-12$\,mag, these correspond to $\lesssim1\%$ of SNe~IIn, which may suggest a more extreme mass-loss event(s) for SN~2021qqp.
By integrating the $r$-band specific luminosity (Figure~\ref{fig:simobj}), the radiated energy during the precursor phase can be roughly estimated to be $8.0\times10^{48}$\,erg.

After the gradual rise, the light curve transitions to a sharp rise ($\approx -30$\,mmag\,day$^{-1}$ in the $g$ and $r$ bands) after $\approx -70$\,days. We consider this transition to be the first light of the SN. With a criterion that the light curve exhibits a monotonic rise above the precursor levels in all bands, we determine the time of SN first light to be $\approx -65\pm5$\,days.  Following the SN first light, the sharp rise continues to $\approx -13$\,days and then transitions to a much sharper rise ($\approx -200$\,mmag\,day$^{-1}$ in the $g$ and $r$ bands) until reaching a maximum of $M_g=-19.4$ and $M_r=-19.5$\,mag. The resulting concave-up curvature of the light curve is atypical of diffusion-dominated light curves (e.g., \citealt{Arnett1980ApJ...237..541A,Arnett1982ApJ...253..785A}). 
The decline from the maximum light is also characterized by a concave-up curvature, albeit with changes in the slope, i.e., a short ($\approx 6$\,day) plateau during $\approx6-12$ days after maximum (see also Figure~\ref{fig:simobj} for an enlarged view around the peak), and roughly a three times longer timescale than the rapidly rising part. 

No photometric measurements are available around $150-220$ days after maximum due to the Sun constraint.
At $\approx 335$\,days, the light curve shows another luminous sharp peak, with $M_g=-16.7$ and $M_r=-17.3$\,mag, with possible bumps preceding at $\approx 240$\,days and following at $\gtrsim 450$\,days (during the current Sun constraint). Unlike the first maximum, this second peak is characterized by a concave-down curvature. The rise starts at $\approx 315$\,days from $M_g=-15.5$ and $M_r=-16.1$\,mag, and the decline lasts until $\approx 380$\,days to $M_g=-13.7$ and $M_r=-15.0$\,mag before transitioning to yet another potential rise. The photometric monitoring is planned to be continued after the current Sun constraint (until 2023 mid-May) to capture further evolution, if any.

Throughout the evolution, the $g-r$ color follows the light curve in that it reaches local minima at the the light-curve peaks. During $-65$ to $-13$\,days, the $g-r$ color stays roughly constant at $0.32$\,mag, albeit with the large scatters. It reaches $-0.07$\,mag at the first light-curve maximum, then becomes redder during the light-curve decline to $0.71$\,mag until $\approx60$\,days and stays roughly constant thereafter. During the second light-curve peak, it becomes bluer again to $0.47$\,mag, then redder to $0.91$\,mag until $\approx380$\,days and stays roughly constant thereafter. Assuming a blackbody SED, the $g-r$ color evolution corresponds to effective temperature evolution of $7000 \rightarrow 10700 \rightarrow 5300\,{\rm K}$ and $5300 \rightarrow 6300 \rightarrow 4700\,{\rm K}$ during the first and second peaks, respectively. The actual effective temperatures are likely higher given the H$\alpha$ line contribution in the $r$-band photometry.

To extract SN and CSM properties from the light-curve modeling in \S\ref{sec:mod}, we construct a bolometric light curve of SN~2021qqp (Figure~\ref{fig:evolution}) by fitting and integrating a blackbody SED to every epoch of photometry containing at least three filters obtained within 2 days of each other. We note that due to the strong H$\alpha$ emission feature (Figure~\ref{fig:spec}), the fitted blackbody temperatures may be underestimated by up to $1500$\,K compared to fits of the spectra (excluding the H$\alpha$ region) at similar epochs; however, the radii are also overestimated such that the resultant bolometric luminosities agree within their error bars. By integrating the bolometric light curve, the total radiated energy is estimated to be $9.5\times10^{49}$ erg, requiring a radiative efficiency of $\sim10\%$ for a typical SN explosion energy of $10^{51}$ erg.
This is divided to $7.2\times10^{49}$ and $2.3\times10^{49}$\,erg at the first and second peaks, respectively.

\subsubsection{Comparison to Other Transients}

\begin{figure*}
    \centering
    \includegraphics[width=1.0\textwidth]{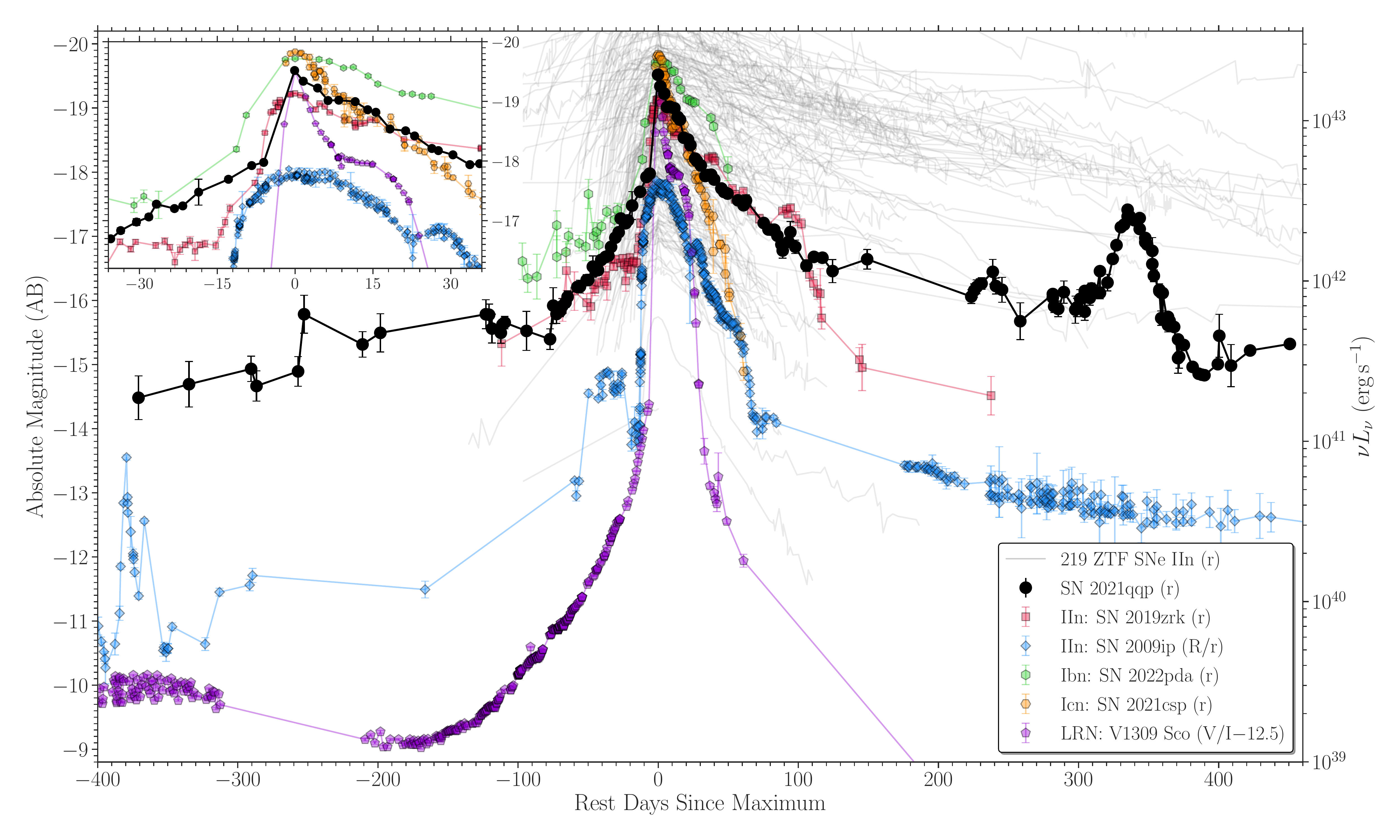}
\caption{Comparison of the $r$/$R$-band light curve of SN~2021qqp with SNe~IIn 2019zrk and 2009ip, Ibn 2022pda, Icn 2021csp, and ZTF SN~IIn sample, as well as LRN V1309~Scorpii (in the $V$/$I$ band shifted by $-12.5$\,mag). SN~2021qqp is characterized by the sharp first peak (half-maximum rise time of $t_{1/2,{\rm rise}}\approx 4$\,days) and distinct luminous second peak ($\approx -17.3$\,mag at $335$\,days). The precursor and peak magnitudes are similar to SNe~2019zrk and 2022pda and brighter than SN~2009ip and LRN V1309~Scorpii, while the decline rate in the $30$\,days after maximum is higher than SNe~2019zrk and 2022pda and lower than SN~2021csp. The objects with precursor events show changes in the slope, i.e., a short ($\approx 5-10$\,days) plateau/peak, within $30$ days of the maximum. Data sources: SNe 2019zrk ($r$ band from \citealt{Fransson2022A&A...666A..79F}), 2009ip ($R$ band from \citealt{Prieto2013ApJ...763L..27P,Mauerhan2013MNRAS.430.1801M,Pastorello2013ApJ...767....1P,Fraser2013MNRAS.433.1312F,Margutti2014ApJ...780...21M,Graham2014ApJ...787..163G}, and $r$ band from \citealt{Graham2014ApJ...787..163G,Graham2017MNRAS.469.1559G}), 2022pda ($r$ band retrieved via ZTF forced-photometry server in this work), and 2021csp ($r$ band from \citealt{Perley2022ApJ...927..180P,Pellegrino2022ApJ...938...73P}), LRN V1309~Scorpii ($I$ band from \citealt{Tylenda2011A&A...528A.114T} and $V$ band from \citealt{Pojmanski2002AcA....52..397P} retrieved via AAVSO International Database), and ZTF SN~IIn sample ($r$ band retrieved via the ALeRCE ZTF Explorer; \citealt{Forster2021AJ....161..242F}).
}
    \label{fig:simobj}
\end{figure*}

In Figure~\ref{fig:simobj}, we compare the $r$-band light curve of SN~2021qqp with several other well-observed interaction-dominated transients: precursor-associated SNe~IIn 2009ip \citep{Smith2010AJ....139.1451S,Foley2011ApJ...732...32F,Prieto2013ApJ...763L..27P, Mauerhan2013MNRAS.430.1801M,Pastorello2013ApJ...767....1P,Fraser2013MNRAS.433.1312F,Margutti2014ApJ...780...21M,Levesque2014AJ....147...23L,Smith2014MNRAS.438.1191S,Graham2014ApJ...787..163G,Mauerhan2014MNRAS.442.1166M,Martin2015AJ....149....9M,Fraser2015MNRAS.453.3886F,Graham2017MNRAS.469.1559G,Reilly2017MNRAS.470.1491R,Smith2022MNRAS.515...71S} and Ibn 2022pda\footnote{Pre-explosion activity is first noted by \cite{Fulton2022TNSAN.198....1F}.} and fast-evolving SN~Icn 2021csp \citep{Fraser2021arXiv210807278F,Perley2022ApJ...927..180P,Pellegrino2022ApJ...938...73P}, as well as the luminous red nova (LRN) V1309 Scorpii, argued to arise from a stellar merger \citep{Mason2010A&A...516A.108M,Tylenda2011A&A...528A.114T}. 

The average precursor and peak magnitudes of SN~2021qqp are more luminous than SN~IIn 2009ip (by $-2.8$ and $-1.6$\,mag, respectively) and LRN V1309 Scorpii (by $-12.4$ and $-12.5$\,mag, respectively), and comparable to SNe~IIn 2019zrk and Ibn 2022pda. The characteristic sharp concave-up curvature around maximum is similar to LRN V1309 Scorpii, albeit with a longer timescale. The peak magnitudes and half-maximum rise time (the duration above the half-maximum on the rising phase) of $t_{1/2,{\rm rise}}\approx 4$\,days (similar in the $g$ band as well) are within the fast blue optical transient (FBOT) regime, mainly composed of SNe~Ibn/Icn and AT 2018cow-like transients (e.g., \citealt{Drout2014ApJ...794...23D,Ho2019ApJ...887..169H,Ho2023ApJ...949..120H,Perley2022ApJ...927..180P,Pellegrino2022ApJ...938...73P}), while the decline rate in the $30$\,days after the maximum is slower than SN~Icn 2021csp and faster than SNe~IIn 2019zrk and Ibn 2022pda. Interestingly, the events with precursors show a short ($\sim 5-10$\,day) plateau/peak within $30$ days of the maximum (see also \citealt{Reguitti2022A&A...662L..10R} for a similar plateau seen in precursor-associated SN~IIn/Ibn 2021foa).
These overall light-curve similarities among different types of transients may suggest a similar progenitor scenario with differing CSM H/He abundance and explosion energy (see further \S\ref{sec:sum}).

We also collect the $r$-band light curves of ZTF objects classified as ``SN~IIn", ``SN~IIn-pec", or ``SLSN-II" on TNS and/or the Weizmann Interactive Supernova Data Repository\footnote{\url{https://wiserep.weizmann.ac.il}} (WISeREP; \citealt{Yaron2012PASP..124..668Y}) using the ALeRCE ZTF Explorer\footnote{\url{https://alerce.online/}} \citep{Forster2021AJ....161..242F} and show them in Figure~\ref{fig:simobj}. Initial visual inspections of the light curves suggest there may be a few more objects with a concave-up curvature like SN~2021qqp, making up only a few percent of the SN~IIn sample. Given their luminous sharp peaks, they may appear as FBOTs in magnitude-limited surveys if they happen at a large distance with only near-peak coverage ($\lesssim-18$\,mag). Among the SN~IIn sample, SN~2021qqp is unique in its distinct sharp second peak. 
A more quantitative sample analysis of the light-curve curvatures will be presented in a future work (D. Hiramatsu et al. in preparation).

\subsection{Spectral Evolution} 
\label{sec:specevo}

As seen in Figure~\ref{fig:spec}, the spectra of SN~2021qqp are initially dominated by Balmer lines on top of a blue continuum ($\leq-13.5$\,days), with weaker He~{\sc i}, Na~{\sc i}, Ca~{\sc i}, and Fe~{\sc ii} lines appearing later as the continuum drops ($\geq314.9$\,days). These spectral behaviors are typically seen in SNe~IIn (e.g., \citealt{Gal-Yam2017hsn..book..195G}). The H$\alpha$ and H$\beta$ lines track the light-curve evolution (Figure~\ref{fig:lc}) in that their luminosities increase at the light-curve peaks, likely indicating interaction with a denser CSM (e.g., \citealt{Chugai1991MNRAS.250..513C,Salamanca1998MNRAS.300L..17S}).

\begin{figure}
    \centering
    \includegraphics[width=0.48\textwidth]{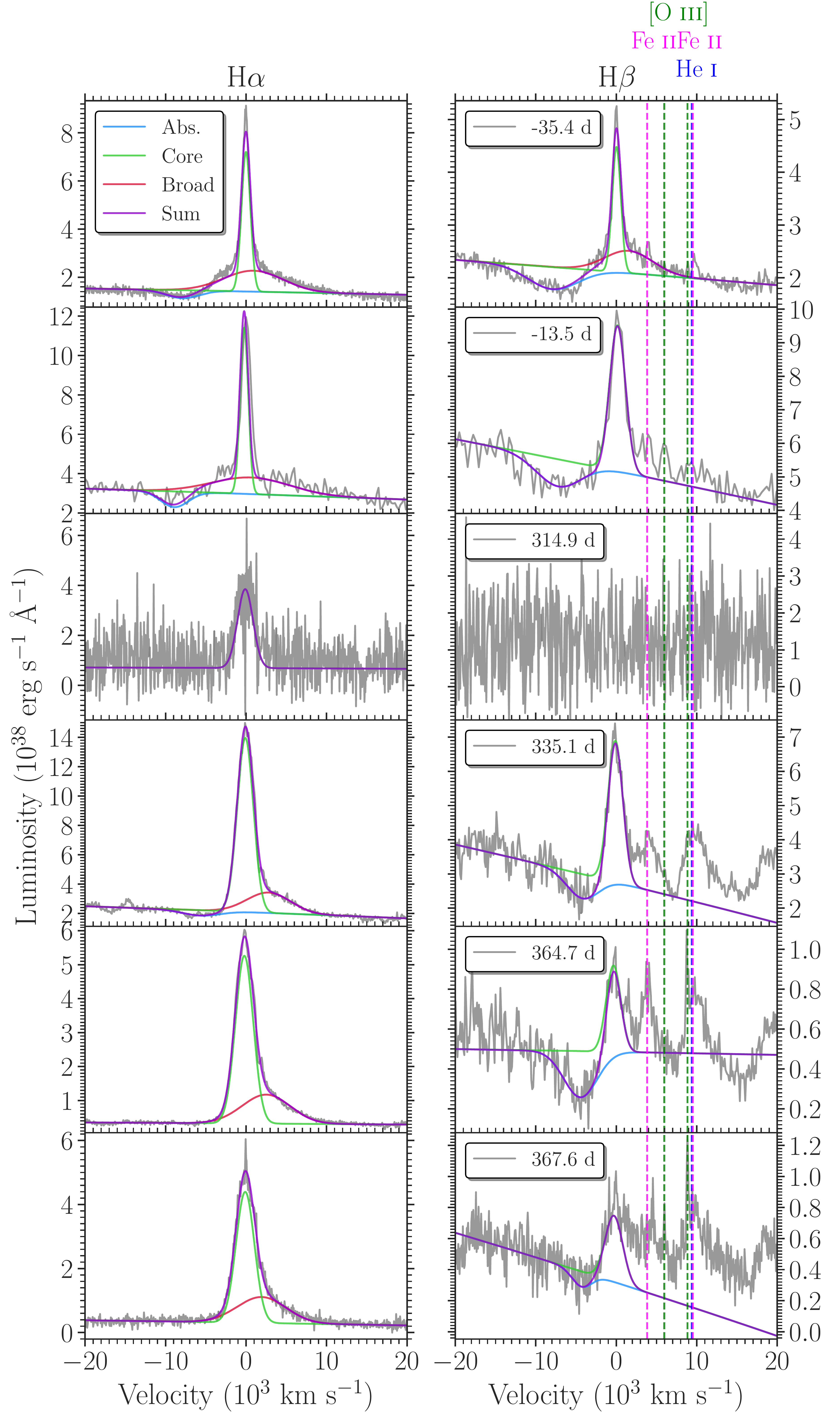}
\caption{Line profile evolution of H$\alpha$ ({\it Left}) and H$\beta$ ({\it Right}) of SN~2021qqp. Multi-component (absorption, core, and broad) Gaussian fits are shown where the absorption minima and core FWHMs are used to estimate the SN--CSM shell and CSM velocites, respectively. The H$\alpha$ and H$\beta$ lines co-evolve with the light curve (Figure~\ref{fig:lc}) in that the more luminous line emission is seen at the light-curve peaks.
}
    \label{fig:HaHb}
\end{figure}

In order to decompose the H$\alpha$ and H$\beta$ line profiles, we fit multi-component Gaussians (absorption, core, and broad; Figure~\ref{fig:HaHb}) when a certain component is visible in a spectrum. In the first two spectra taken during the rise to the light-curve maximum ($-35.4$ to $-13.5$\,days), the H$\alpha$ and H$\beta$ line profiles can be fit well with all three components,\footnote{Except for the broad H$\beta$ component in the second spectrum due to its low signal-to-noise ratio, which may result in an overestimation of the core component.} with the resulting absorption minima at $\approx 7000-8900$\,km\,s$^{-1}$ and core FWHMs of $\approx 1300-1900$\,km\,s$^{-1}$. 
In the fourth spectrum taken at the light-curve second peak ($335.1$\,days),\footnote{We exclude the third spectrum due to its low signal-to-noise ratio.} all three components are still visible in H$\alpha$ with the broad component peak redshifted, while no broad component is visible in H$\beta$ due to the presence of strong Fe~{\sc ii} emission. In the final two spectra taken during the decline from the second peak ($364.7$ to $367.6$\,days), absorption component is still visible in H$\beta$, but not clearly in H$\alpha$. During the second peak ($314.9$ to $367.6$\,days), the H$\alpha$ and H$\beta$ absorption minima and core FWHMs correspond to $\approx4200-5600$ and $\approx2100-2680$\,km\,s$^{-1}$, respectively.

We associate the H$\alpha$ absorption and line core velocities with the SN--CSM shell and CSM, respectively (the core CSM component and the absorption + broad P Cygni components from the SN--CSM shell; Figure~\ref{fig:evolution}) and reproduce them with the light-curve modeling in \S\ref{sec:mod}.
The SN--CSM shell velocity decreases from $\approx8500$ to $\approx5600$\,km\,s$^{-1}$ from the first to second light-curve peaks, while the CSM velocity increases from $\approx1300$ to $\approx2300-2680$\,km\,s$^{-1}$ from the first to second light-curve peaks (i.e., earlier CSM ejection is moving faster). This increasing CSM velocity is on the fast end of those seen in typical SN precursors \citep{Ofek2014ApJ...789..104O,Strotjohann2021ApJ...907...99S} and comparable to some faster components seen, for example, in the SN~2009ip precursors \citep{Smith2010AJ....139.1451S,Foley2011ApJ...732...32F,Mauerhan2013MNRAS.430.1801M, Pastorello2013ApJ...767....1P} and some giant eruptions of Eta Carinae \citep{Smith2008Natur.455..201S,Smith2018MNRAS.480.1457S}.
Finally, we note that a narrower wind P Cygni component is not detected on top of the core component, but we cannot rule out its existence below our spectral resolution ($\lesssim200$\,km\,s$^{-1}$).

\section{Modeling with CSM Interaction} 
\label{sec:mod}

\subsection{Analytical Model}

Given the observed properties of SN~2021qqp (\S\ref{sec:ana}), we model the light curve assuming the emission is produced purely by the shock interaction between the SN ejecta and CSM. For multi-peak transients like SN~2021qqp, the CSM is expected to have a more complicated profile than a single power law as adopted in many previous studies (e.g., \citealt{Chatzopoulos+2012,Moriya+2013d}). While it is still possible to use a more complicated functional form for the CSM \citep[e.g., a single power law and Gaussian;][]{Gomez+2021,Hosseinzadeh+2022}, here we propose a more flexible approach that does not require the functional form of the CSM profile. A more detailed description and application to other interacting transients will be presented in a forthcoming paper (T.~Matsumoto et al.~in preparation).

We consider SN ejecta colliding with the CSM, which forms a shell separated from the un-shocked CSM and SN ejecta by a forward shock (FS) and a reverse shock (RS), respectively. Assuming the shell is geometrically thin and hence its location, velocity, and mass are given by $\Rsh$, $\vsh$, and $\Msh$, the time evolution of the shell is described by two equations \citep{Chevalier1982b,Moriya+2013d},

\begin{align}
\Msh\frac{d\vsh}{dt}&=4\pi \Rsh^2\rho_{\rm SN}(\vsn-\vsh)^2-4\pi \Rsh^2 \rho_{\rm CSM}(\vsh-\vcsm)^2 ,
	\label{eq:dvdt}\\
\frac{dM_{\rm sh}}{dt}&=4\pi \Rsh^2\rho_{\rm SN}(\vsn-\vsh)+4\pi \Rsh^2\rho_{\rm CSM}(\vsh-\vcsm)\ ,
	\label{eq:dmdt}
\end{align}

\noindent where $\rho_{\rm SN}$ and $\vsn=\Rsh/t$ are the density and velocity of the un-shocked SN ejecta at $\Rsh$, respectively, and $\rho_{\rm CSM}$ and $\vcsm$ are the un-shocked CSM density and velocity, respectively. We assume that the un-shocked SN ejecta expands homologously, and its density is approximated by a broken power-law profile \citep[e.g.,][]{Chevalier&Fransson1994,Matzner&McKee1999},

\begin{align}
\rho_{\rm SN}(v,t)=A\begin{cases}
(v/v_*)^{-\delta}& \,\,v<v_*\ ,\\
(v/v_*)^{-n}&\,\,v\ge v_*\ ,
\end{cases}
	\label{eq:rho_sn}
\end{align}

\noindent where 

\begin{align}
v_*&=\sqrt{\frac{2(5-\delta)(n-5)E_{\rm SN}}{(3-\delta)(n-3)M_{\rm SN}}}
	\label{eq:v*}\\
&\overset{\substack{\delta=0\\n=12}}{\approx }3600{\,\rm km\,s^{-1}\,}\left(\frac{M_{\rm SN}}{10\,\Msun}\right)^{-1/2}\left(\frac{E_{\rm SN}}{10^{51}{\,\rm erg}}\right)^{1/2}\ ,
	\nonumber
\end{align}

\noindent where $M_{\rm SN}$ and $E_{\rm SN}$ are the total  mass and kinetic energy of the SN ejecta, respectively. Typically, $\delta=0-1$ and $n\simeq12$ is expected for red supergiant progenitors or $n\simeq10$ for progenitors with radiative envelopes, e.g., blue supergiants \citep{Matzner&McKee1999}. The normalization $A$ is given such that the integration of $\rho_{\rm SN}$ gives $M_{\rm SN}$.

In contrast to previous works that used a parameterized CSM profile, we determine it by requiring the shock luminosity to produce the observed bolometric luminosity,

\begin{align}
L_{\rm obs}\approx \varepsilon_{\rm FS} L_{\rm kin,FS}+\varepsilon_{\rm RS} L_{\rm kin,RS},
	\label{eq:Lobs}
\end{align}
where the dissipated kinetic energy luminosities at the FS and RS are given by
\begin{align}
L_{\rm kin,FS}&=\frac{9\pi}{8} \Rsh^2 \rho_{\rm CSM}(\vsh-\vcsm)^3\ ,
	\label{eq:Lfs}\\
L_{\rm kin,RS}&=\frac{9\pi}{8}\Rsh^2 \rho_{\rm SN}(\vsn-\vsh)^3\ ,
	\label{eq:Lrs}
\end{align}

\noindent where we assumed an ideal gas \citep[see, e.g.,][]{Metzger+2014b}. The quantities $\varepsilon_{\rm FS}$ and $\varepsilon_{\rm RS}$ represent the conversion efficiency from the dissipated energy to optical photons. While the efficiencies may vary with time \citep[e.g.,][]{Tsuna+2019}, we simply assume a constant and identical efficiency for both FS and RS: $\varepsilon_{\rm FS}=\varepsilon_{\rm RS}=\varepsilon$. With Equations~\eqref{eq:Lobs} and \eqref{eq:Lfs}, the CSM density is estimated by

\begin{align}
\rho_{\rm CSM}=\frac{8(L_{\rm obs}-\varepsilon L_{\rm kin,RS})}{9\pi \varepsilon\Rsh^2(\vsh-\vcsm)^3}\ .
\end{align}

\noindent We can then rewrite Equations~\eqref{eq:dvdt} and \eqref{eq:dmdt} without the CSM density:

\begin{align}
\Msh\frac{d\vsh}{dt}
&=-\frac{32L_{\rm obs}}{9\varepsilon(\vsh-\vcsm)}+\frac{32(\vsn-\vcsm)L_{\rm kin,RS}}{9(\vsn-\vsh)(\vsh-\vcsm)}\ ,
	\label{eq:dvdt2}\\
\frac{dM_{\rm sh}}{dt}
&=\frac{32L_{\rm obs}}{9\varepsilon(\vsh-\vcsm)^2}
	\nonumber\\
+&\frac{32(\vsn-\vcsm)(2\vsh-\vsn-\vcsm)L_{\rm kin,RS}}{9(\vsn-\vsh)^2(\vsh-\vcsm)^2}\ .
	\label{eq:dmdt2}
\end{align}

\noindent We note that the kinetic luminosity, $L_{\rm kin,RS}$, can be calculated for a given $R_{\rm sh}$ and $\vsh$ for an assumed SN ejecta profile by using Equation~\eqref{eq:Lrs}.

Equations~\eqref{eq:dvdt2} and \eqref{eq:dmdt2}, with $d\Rsh/dt=\vsh$, can be solved for a given observed bolometric light curve ($L_{\rm obs}$) and assumed SN properties ($E_{\rm SN}$, $M_{\rm SN}$, $\delta$, and $n$), the emission efficiency $\varepsilon$, and the CSM velocity $\vcsm$. As an initial condition, we assume that the interaction happens at $t_0$ since the SN explosion (at $t_{\rm exp}$) with the initial shell velocity $v_{\rm sh,0}$. The initial shell mass is dominated by the swept-up SN ejecta, which is given by $M_{\rm sh,0}=M_{\rm SN}(>v_{\rm sh,0})=\int_{v_{\rm sh,0}}4\pi r^2\rho_{\rm SN}dr$. Once the time evolution of $\Rsh$ and $\vsh$ are obtained, the density profile is reconstructed using Equation~\eqref{eq:Lfs}. We assume the shock power completely dominates the light curve and neglect any radioactive nickel heating. This can be justified for SN~2021qqp given its sharp concave-up first peak and blueward color evolution (Figure~\ref{fig:lc}), indicating that shock interaction shapes the light curve.

\subsection{Application to SN\,2021qqp}

\begin{figure*}
\centering
\includegraphics[width=1\textwidth]{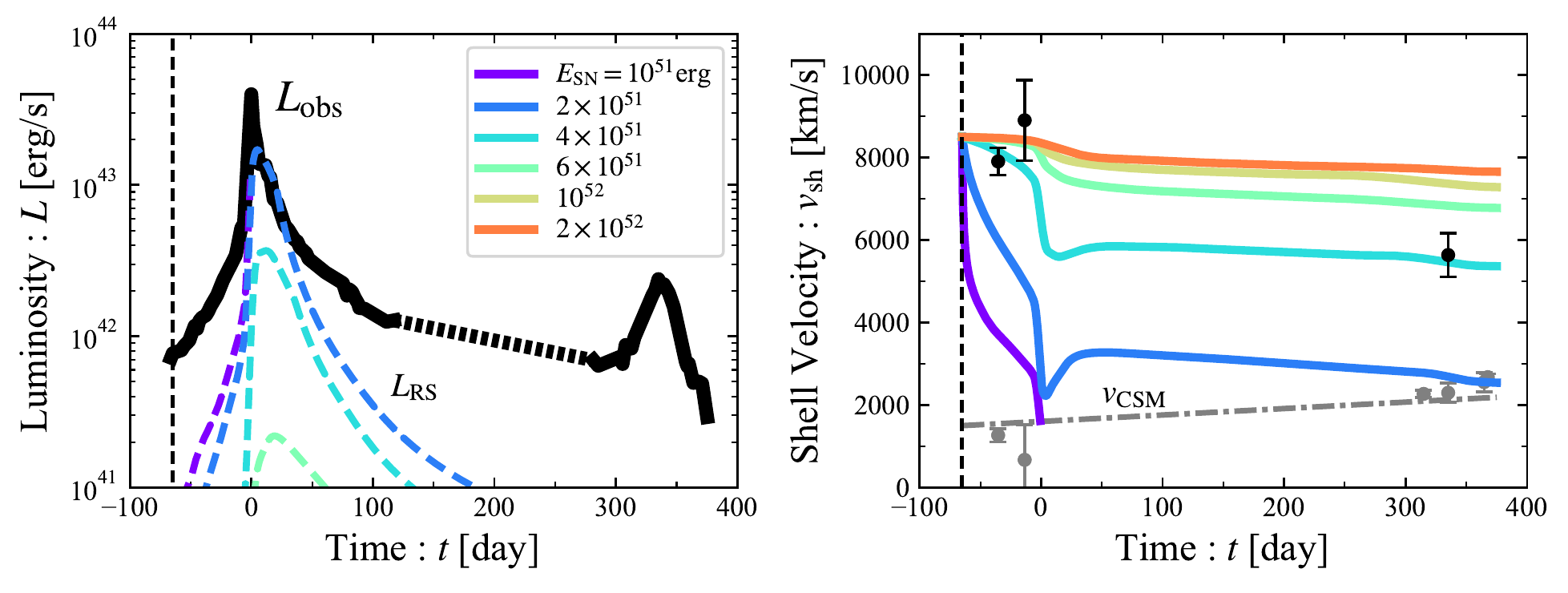}
\caption{The bolometric light curve ({\it Left}) and time evolution of shell velocity ({\it Right}) for different SN explosion energies, with other parameters fixed to $M_{\rm SN}=10\,\Msun$, $n=12$, $\delta=0$, $\varepsilon=0.3$, $v_{\rm sh,0}=8500\,\rm km\,s^{-1}$, and $t_{\rm exp}=-65\,\rm days$ (shown by the vertical black dashed line). We assume that the shock interaction starts almost at the same time as the SN explosion ($t_0-t_{\rm exp}=0.01\,\rm day$). The velocities inferred from the absorption and core-emission of the H$\alpha$ lines (Figure~\ref{fig:HaHb}) are shown by black and gray points, respectively. The former and latter likely correspond to the shell and CSM velocities, respectively. The CSM velocity increases with time (gray dashed-dotted line).
In the left panel, the dashed curves show the RS luminosities.  For this particular set of assumed parameters, we find $E_{\rm SN}\approx 4\times 10^{51}$ erg matches both the velocity evolution and the requirement that $L_{\rm RS}\lesssim L_{\rm obs}$. In the left panel, the linearly interpolated gap in the observed light curve for $100$--$300\,\rm days$ is shown by a dotted line.}
\label{fig:evolution}
\includegraphics[width=1\textwidth]{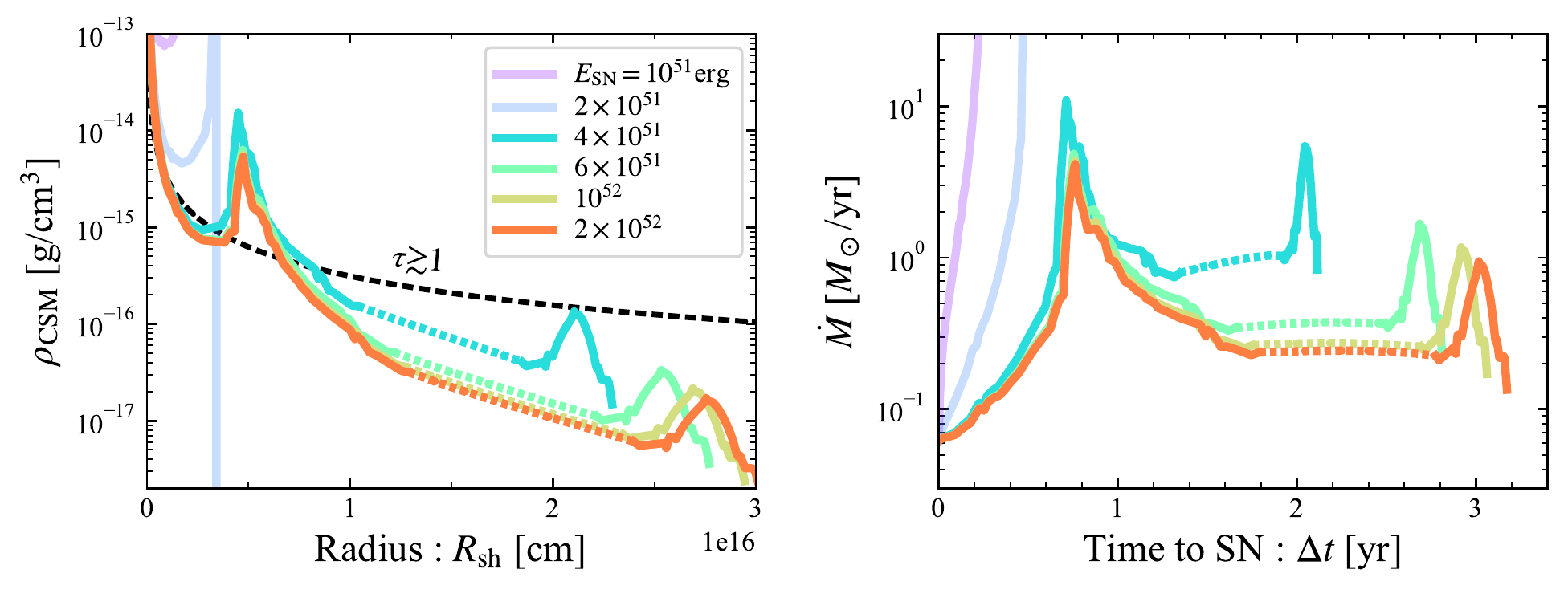}
\caption{CSM density at the FS crossing ({\it Left}) and time evolution of the mass-loss rate relative to the time of SN explosion ({\it Right}) reconstructed by using the results in Figure~\ref{fig:evolution}. Both CSM density and mass-loss rate have a double peak as expected from the observed light curve. The SN energies of $E_{\rm SN}\leq2\times10^{51}\,\rm erg$ (shown with light colors) are rejected because they give a stalling shell or unphysical negative FS luminosity (i.e., $L_{\rm RS} \gtrsim L_{\rm obs}$). In the left panel, the black dashed curve shows a rough estimate of the optical depth ($\rho=1/(\kappa \Rsh)$ with $\kappa=0.32\,\rm cm^2\,g^{-1}$). The CSM density should roughly satisfy $\tau\gtrsim1$ so that we observe thermal (optical) emission. The dotted segment of each result corresponds to the gap in the light curve for $100-300\,\rm days$ (see the left panel of Figure~\ref{fig:evolution}).
}
\label{fig:profile}
\end{figure*}

We apply our analytical model to SN 2021qqp to determine the required SN and CSM properties. In Figure~\ref{fig:evolution}, we show representative solutions for different assumed SN energies and fixed parameters of $M_{\rm SN}=10\,\Msun$, $\delta=0$, $n=12$, and $\varepsilon=0.3$. The choice of the values of $\delta$ and $n$ does not noticeably affect the result. The value of $\varepsilon$ is motivated by having a mildly optically thick CSM, as well as the required energetics (see below). We assume that the SN explosion happened at a phase of $t_{\rm exp}=-65 \,\rm days$ (i.e., at the SN first light; \S\ref{sec:lc}) and the shock interaction started $0.01\,\rm days$ after the explosion (i.e., $t_0-t_{\rm exp}=0.01\,\rm day$). The following results do not change significantly for different values of $t_{\rm exp}$ and $t_0$ unless it is after the first peak (i.e., $t_0\gtrsim 0\,\rm day$). Motivated by the observed H$\alpha$ line profiles (Figure~\ref{fig:HaHb}), the initial shell velocity is set to $v_{\rm sh,0}=8500\,\rm km\,s^{-1}$, and we consider a gradually increasing CSM velocity from $\vcsm\simeq1500$ to $2200\,\rm km\,s^{-1}$. As the observed bolometric light curve has a gap between 100 and 300\,days due to the Sun constraint, we linearly interpolate the light curve to fill the gap.

The right panel of Figure~\ref{fig:evolution} shows the time evolution of the shell velocity. At the fist peak, the shell decelerates by colliding with the massive CSM, producing the first light-curve peak. For large SN energies, the deceleration is weak, and the shell moves almost at a constant velocity, while for smaller energies, the deceleration is significant and even stalls the shell. These evolution for very low and high $E_{\rm SN}$ are inconsistent with the observed line velocity (black points). A mild deceleration, as required by the spectroscopic data, is realized only for a moderate SN energy of $E_{\rm SN}\approx 4\times10^{51}\,\rm erg$. The left panel of Figure~\ref{fig:evolution} depicts the resulting RS shock luminosity (dashed curves), as well as the observed bolometric luminosity (black curve). By construction, the observed luminosity is automatically reproduced by the sum of the FS and RS luminosities. When the shell decelerates at the first peak, the SN ejecta catches up with the shell and powers bright RS emission. In particular, for drastic deceleration, the RS luminosity exceeds the observed bolometric luminosity, and such a solution ($E_{\rm SN}\lesssim 2\times10^{51}\,\rm erg$) can be rejected.
For this particular choice of $M_{\rm SN}=10\,M_\odot$, the modeled RS luminosity and velocity evolution for $E_{\rm SN}\approx 4\times10^{51}\,\rm erg$ are both consistent with SN~2021qqp.

In Figure~\ref{fig:profile}, we show the CSM density when the FS arrives at each radius and the reconstructed mass-loss rates resulting from the models in Figure~\ref{fig:evolution}. It should be noted that the CSM density does not represent the CSM profile because the CSM expands at different velocities. For a less energetic SN, the CSM density is higher to compensate for the lower shell velocity and still reproduce the observed luminosity. For $\varepsilon=0.3$, the CSM has a moderate optical depth $\tau\sim 1$, which may be consistent with the observed optical emission.\footnote{It is not trivial whether the CSM optical depth should be smaller or larger than unity. On the one hand, the optical depth may be required to be $\tau>1$ to avoid bright hard X-ray emission. On the other hand, $\tau<1$ may also be required to explain the observed velocities in the Balmer lines. These disagreeing requirements likely indicate the limitation of the assumption of spherical symmetry. We defer a more detailed discussion to a future work.} Corresponding to the double-peaked light curve, we find that the CSM density has two distinct peaks, indicating that the progenitor experienced two distinct mass-loss episodes. To explore this structure, we translated the density to a mass-loss rate in the right panel of Figure~\ref{fig:profile}. We find that the mass-loss rate is as high as $\sim 1-10\,\Msun\,\rm yr^{-1}$ at $\approx 0.8$ and $\approx 2$\,yr before the explosion (for the model with $E_{\rm SN}\approx 4\times 10^{51}$ erg) over relatively short episodes of $\approx 0.2-0.5\,\rm yr$. The mass-loss episode directly preceding the SN explosion is more extreme. The total CSM mass is $M_{\rm CSM}\approx 2.7\,\Msun$ (for $E_{\rm SN}=4\times10^{51}\,\rm erg$). The bluer $g-r$ color evolution and more luminous H$\alpha$ and H$\beta$ line emission around the light-curve peaks can also be explained by the interaction with these denser CSM.

\begin{figure}
\centering
\includegraphics[width=0.5\textwidth]{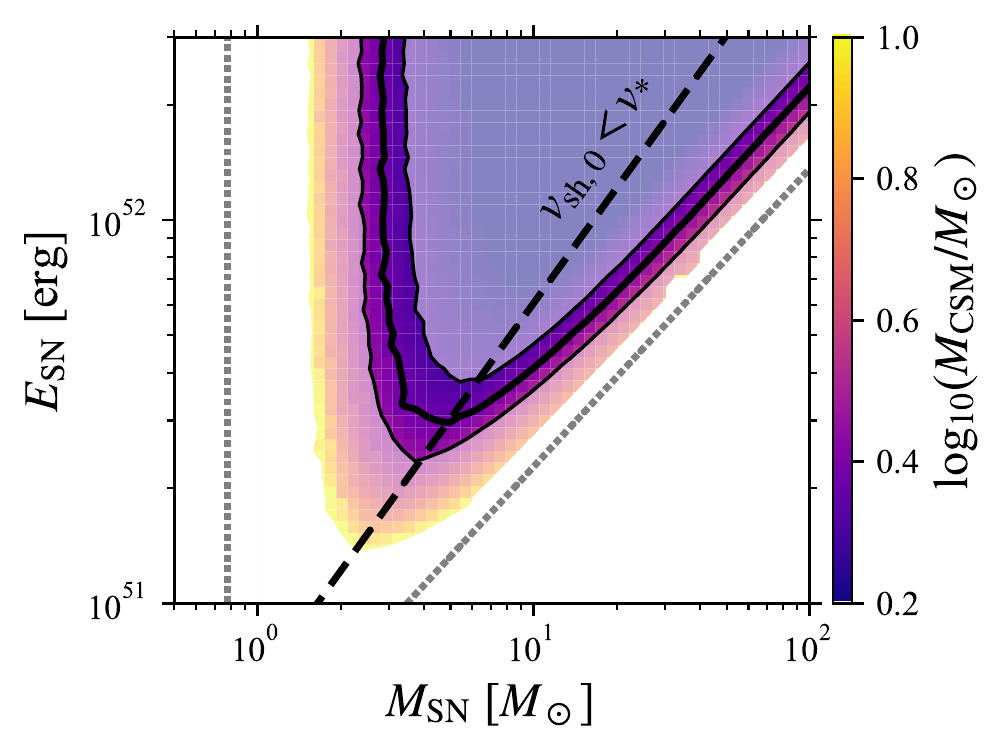}
\caption{Estimated CSM mass for different SN ejecta masses and energies. The other parameters are the same as those adopted for Figure~\ref{fig:evolution}. The parameters giving the shell velocity consistent with the observed one ($5640\pm 530\,\rm km\,s^{-1}$ at $335$\,days) are shown by black curves.
Along the black dashed line, the initial velocity is the same as the characteristic ejecta velocity (Equation~\ref{eq:v*}). The dotted gray vertical and diagonal lines (Equations~\ref{eq:Mx} and \ref{eq:Ex}, respectively) give rough boundaries within which the shell does not stall during the observation. 
}
\label{fig:parameter}
\end{figure}

The results above were provided for a fixed example value of $M_{\rm SN}=10\,\Msun$. To explore the parameter space of SN properties more broadly, we carried out the same analysis for different SN ejecta masses and energies to find the parameter space consistent with the observed light curve and velocities. In Figure~\ref{fig:parameter}, we show the allowed parameter region. The colored region denotes the space over which the shell expands continuously without stalling and gives a finite $M_{\rm CSM}$. We derive the parameter space satisfying the condition that the shell decelerated mildly and its velocity at $335$\,days is consistent with the observed value of $5640\pm530\,\rm km\,s^{-1}$. The allowed region is enclosed by the black thin curves accounting for the velocity uncertainty, while the black thick curve corresponds to the velocity being exactly the same as the observed value. Along the allowed region, the CSM mass is relatively well constrained to $M_{\rm CSM}\approx 2-4\,\Msun$. The SN ejecta is constrained to have an energy of $\gtrsim 3\times10^{51}\,\rm erg$, which is slightly larger than typical values but still consistent with the stellar explosion scenario, potentially further enhanced by jets (e.g., \citealt{Soker2010MNRAS.401.2793S,Papish2011MNRAS.416.1697P,Shishkin2023MNRAS.522..438S}).

The allowed parameter space exhibits two branches, based on the initial shell velocity $v_{\rm sh,0}$. Too-low initial velocity $v_{\rm sh,0}<v_*$ (left of the black dashed diagonal line in Figure~\ref{fig:parameter}) means that most of the SN ejecta forms a shell instantaneously when the shock interaction starts, which is not natural, and we therefore disfavor this portion of the parameter space. More natural solutions appear for the initial shell velocity larger than $v_*$, which means that the shock interaction begins at the high-velocity tail in the SN ejecta. In this case, the initial shell mass is much smaller than the whole SN ejecta, and the shell readily decelerates when it collides with a dense CSM bump. 

We can derive critical conditions for which the shell stalls during the observation. These conditions are obtained by considering the initial deceleration timescale of the shell,

\begin{align}
t_{\rm dec}&=\frac{v_{\rm sh,0}}{d\vsh/dt}\approx \frac{9\varepsilon M_{\rm sh,0}v_{\rm sh,0}^2}{32L_{\rm obs}}\ ,
\end{align}

\noindent where we used Equation~\eqref{eq:dvdt2} neglecting the CSM velocity and RS luminosity. When the initial shell velocity is smaller than the SN characteristic velocity, $v_*$, the deceleration timescale is determined by the SN mass. By equating $t_{\rm dec}$ with the characteristic emission timescale (e.g., peak timescale), we have a critical mass below which the shell stalls over the emission timescale,

\begin{align}
M_{\rm SN,\times}\approx 0.6{\,\Msun\,}\left(\frac{\varepsilon}{0.3}\right)^{-1}\left(\frac{E_{\rm rad}}{10^{50}\,\rm erg}\right)\left(\frac{v_{\rm sh,0}}{10^4\,\rm km\,s^{-1}}\right)^{-2}\ ,
    \label{eq:Mx}
\end{align}

\noindent where $E_{\rm rad}$ is the radiated energy over the peak timescale. The gray dotted vertical line shows this condition.

For the case of higher initial velocity $v_{\rm sh,0}>v_*$, the shell's mass is smaller than the total SN ejecta mass. In the same way as for Equation \eqref{eq:Mx}, we have a relation between $E_{\rm SN}$ and $M_{\rm SN}$ corresponding to the gray dotted diagonal line,

\begin{align}
E_{\rm SN,\times}>&2.9\times10^{51}{\,\rm erg\,}\left(\frac{\varepsilon}{0.3}\right)^{-2/9}\left(\frac{E_{\rm rad}}{5\times10^{48}\,\rm erg}\right)^{2/9}
	\nonumber\\
&\left(\frac{M_{\rm SN}}{10\,\Msun}\right)^{7/9}\left(\frac{v_{\rm sh,0}}{10^4{\,\rm\,km\,s^{-1}}}\right)^{14/9}\ ,
    \label{eq:Ex}
\end{align}

\noindent where we used $n=12$ and $\delta=0$ and the radiated energy up to $-20$\,days because the shell decelerates and stalls roughly before this timescale. This condition also gives a scaling law for the allowed parameter space,

\begin{align}
E_{\rm SN}\approx 4\times10^{51}{\,\rm erg\,}\left(\frac{\varepsilon}{0.3}\right)^{-2/9}\left(\frac{M_{\rm SN}}{10\,\Msun}\right)^{7/9}\ ,
\end{align}

\noindent along the black thick curve with $v_{\rm sh, 0} > v_*$. Within a reasonable energy range of $E_{\rm SN}\approx (3-10)\times10^{51}$\,erg in the allowed parameter space, the corresponding allowed mass range is $M_{\rm SN}\approx 5-30\,M_\odot$.

\section{Summary and Discussion} 
\label{sec:sum}

Before addressing the implications of our findings, below we summarize the key observed and modeled properties of SN~2021qqp (\S\ref{sec:ana} and \S\ref{sec:mod}): 

\begin{itemize}
\vspace{-4pt}
  \item A luminous ($-15.8\,{\rm mag} \lesssim M_{g,r,i}\lesssim -14.5\,{\rm mag}$) and long-lasting ($\sim 300$\,days) precursor leading up to the SN explosion.
\vspace{-4pt}
  \item A multi-peaked SN light curve with a first concave-up peak ($M_g=-19.4$, $M_r=-19.5$\,mag) $\approx 65$ days after explosion and a second concave-down peak ($M_g=-16.7$, $M_r=-17.3$\,mag) $\approx 335$ days after first peak.
\vspace{-4pt}
  \item Bluer $g-r$ colors (i.e., higher temperatures) around the light-curve peaks.
\vspace{-4pt}
  \item Spectra dominated by Balmer lines, with weaker He~{\sc i}, Na~{\sc i}, Ca~{\sc i}, and Fe~{\sc ii} lines.
\vspace{-4pt}
  \item More luminous H$\alpha$ and H$\beta$ line emission around the light-curve peaks.
\vspace{-4pt}
  \item A decreasing shell velocity (from $\approx8500$\,km\,s$^{-1}$ in the first peak to $\approx 5600$\,km\,s$^{-1}$ in the second peak) and increasing CSM velocity (from $\approx1300$\,km\,s$^{-1}$ in the first peak to $\approx 2500$\,km\,s$^{-1}$ in the second peak).
\vspace{-4pt}
  \item Two distinct CSM density peaks from episodic mass loss ($\dot{M}\approx10\,M_\odot\,{\rm yr}^{-1}$ about $0.8$\,yr before explosion and $\approx5\,M_\odot\,{\rm yr}^{-1}$ about $2$\,yr before explosion), with a total CSM mass of $M_{\rm CSM}\approx 2-4\,M_\odot$. 
\vspace{-4pt}
  \item An allowed SN ejecta mass range of $M_{\rm SN}\approx 5-30\,M_\odot$ for an explosion energy range of $E_{\rm SN}\approx (3-10)\times10^{51}$\,erg, satisfying a consistent RS luminosity limit and the observed velocity evolution. 
\vspace{-4pt}
\end{itemize}
 
These observed and modeled properties suggest eruptive mass-loss episodes preceding an energetic explosion. The final mass-loss episode is likely related to the pre-explosion outburst detected starting about a year before the explosion with a luminosity of $\approx 3\times10^{41}\,\rm erg\,s^{-1}$ (Figure~\ref{fig:simobj}). Such a precursor can be produced by an eruption of a giant star ($\sim 10^2\,\Rsun$) with an ejection mass of a few $M_\odot$ and velocity of $\sim10^3$\,km\,s$^{-1}$ \citep{Matsumoto&Metzger2022}, which is consistent with the inferred CSM properties (Figures~\ref{fig:evolution}--\ref{fig:parameter}). The previous mass-loss episode with a less violent mass ejection may result in a luminosity around the detection threshold of $\approx 10^{41}\,\rm erg\,s^{-1}$ with a shorter duration of $\approx 30$\,days. We remark that the ejected material in this episode $2$\,yr before explosion would not affect the observed precursor because its optical depth is at most $\tau\sim1$ (see the left panel in Fig.~\ref{fig:profile}).

We now discuss two possible progenitor channels for generating SN~2021qqp:\footnote{The sharp light-curve morphology also resembles gravitational microlensing events (e.g., \citealt{Gaudi2012ARA&A..50..411G}); however, the chromatic evolution seen in SN~2021qqp excludes such a possibility.} (i) stellar activity preceding an SN explosion (e.g., \citealt{Poelarends2008ApJ...675..614P,Quataert2012MNRAS.423L..92Q,Jones2013ApJ...772..150J,Shiode2014ApJ...780...96S,Quataert2016MNRAS.458.1214Q,Fuller2017MNRAS.470.1642F,Doherty2017PASA...34...56D,Fuller2018MNRAS.476.1853F,Wu2021ApJ...906....3W,Wu2022ApJ...930..119W,Matsumoto&Metzger2022}) and (ii) common envelope (CE) evolution preceding a stellar merger (e.g., \citealt{Chevalier2012ApJ...752L...2C,Pejcha2016MNRAS.455.4351P,Pejcha2017ApJ...850...59P,MacLeod2017ApJ...835..282M,Metzger2017MNRAS.471.3200M,MacLeod2018ApJ...863....5M,MacLeod2018ApJ...868..136M,Schroder2020ApJ...892...13S,MacLeod2020ApJ...893..106M,MacLeod2020ApJ...895...29M,Metzger2022ApJ...932...84M,Matsumoto2022ApJ...938....5M}).

As the first scenario, we consider the eruptive mass-loss episodes from pre-explosion stellar activity. Typical mass-loss rates and velocities from supergiant stars (e.g., extreme red supergiant and super-asymptotic giant branch stars; \citealt{Smith2014ARA&A..52..487S,Smith2017hsn..book..403S}) of $\dot{M}\lesssim10^{-3}\,M_\odot\,{\rm yr}^{-1}$ and $v_{\rm CSM}\leq100$\,km\,s$^{-1}$, respectively, are lower than the inferred CSM values. Enhancements in mass loss may be achieved with thermal pulses and/or internal gravity waves excited by late-stage nuclear burning (e.g., He, C, O/Ne, and Si; \citealt{Poelarends2008ApJ...675..614P, Shiode2014ApJ...780...96S,Fuller2017MNRAS.470.1642F,Wu2022ApJ...930..119W}); however, the expected maximum energy output of $\lesssim10^{48}$\,erg is lower than the observed radiated energy of $\approx 8\times10^{48}$\,erg during the SN~2021qqp precursor.  Instead, LBV giant eruptions, like those seen in Eta Carinae, P Cygni, and SN impostors ($\dot{M}\sim10^{-2}-10\,M_\odot\,{\rm yr}^{-1}$ and $v_{\rm CSM}\sim100-1000$\,km\,s$^{-1}$; e.g., \citealt{Humphreys1994PASP..106.1025H,Davidson1997ARA&A..35....1D,Vink2012ASSL..384..221V,VanDyk2012ASSL..384..249V,Smith2014ARA&A..52..487S,Smith2017RSPTA.37560268S}), may be responsible for the precursor, given their comparable luminosities. Although the high initial mass range of LBVs ($\gtrsim 30\,M_\odot$) seems to be at odds with the estimated SN ejecta mass ($\lesssim30\,M_\odot$ for $E_{\rm SN}\lesssim10^{52}$\,erg; Figure~\ref{fig:parameter}), this might be reconciled through significant mass loss in successive eruptions prior to the explosion (e.g., \citealt{Vink2012ASSL..384..221V,Smith2019MNRAS.488.1760S}).

A more exotic explanation in the context of eruptive mass loss is a PPISN (e.g., \citealt{Woosley2007Natur.450..390W,Blinnikov2010PAN....73..604B,Moriya2013MNRAS.428.1020M,Woosley2017ApJ...836..244W}). The ejecta mass and explosion energy inferred for SN~2021qqp may be reproduced with a PPISN model with an initial mass range of $\sim110-140\,M_\odot$ \citep{Woosley2017ApJ...836..244W}.  In this model, the bulk of the mass is lost during the progenitor's evolution and pair-instability pulses, followed by a final collapse to a black hole (BH). The expected pulsational pulse intervals span a wide range, but if several pulses with H-rich mass ejection could occur in the last few years before collapse, the CSM configuration might resemble that of SN~2021qqp.

On the other hand, the similarity of SN~2021qqp's light curve to that of LRN V1309 Scorpii (Figure~\ref{fig:simobj}) motivates the alternative scenario of a stellar merger. Although the energetics for LRNe from a typical stellar merger ($\lesssim10^{41}$\,erg\,s$^{-1}$; \citealt{Pejcha2016MNRAS.455.4351P,Pejcha2017ApJ...850...59P,Metzger2017MNRAS.471.3200M,Matsumoto2022ApJ...938....5M}) are well below what is required for SN\,2021qqp, a merger of a Wolf-Rayet (WR) star and a neutron star (NS) or BH (e.g., \citealt{Chevalier2012ApJ...752L...2C,Schroder2020ApJ...892...13S,Metzger2022ApJ...932...84M}) may be able to reproduce the light curve of SN~2021qqp. In this scenario, a massive star ($\gtrsim 20\,M_\odot$) and NS/BH (from an earlier SN) undergo CE evolution, leaving a tight WR-NS/BH binary. The H-rich CE ejection is manifested as a precursor, and if a merger-induced explosion ($>10^{51}$\,erg) follows promptly ($\lesssim10$\,yr), the system may resemble a precursor-associated SN~IIn. The estimated CSM and SN ejecta masses for SN~2021qqp are within the model expectations if the merger happens within $\sim 1$\,yr of the CE ejection.  However, it is unclear if this model can produce successive mass ejections with $\sim10^{3}$\,km\,s$^{-1}$ that can reproduce the multi-peaked CSM density profile of SN~2021qqp (Figure~\ref{fig:profile}). This may be possible if several eccentric encounters happen prior to the onset of the CE phase (e.g., \citealt{Vigna-Gomez2020PASA...37...38V,Vick2021MNRAS.503.5569V}), leading to successive CSM peaks as increasing quasi-periodic mass ejections are expected toward the merger (see e.g., \citealt{Soker2013ApJ...764L...6S,Kashi2013MNRAS.436.2484K} for an application to SN~2009ip).  Continued optical monitoring of SN~2021qqp may reveal the presence of even earlier CSM peaks that may be expected in this eccentric encounter scenario.

As proposed in \citet{Metzger2022ApJ...932...84M}, depending on the time delay between the CE ejection and stellar merger, this scenario may be responsible for the light-curve similarities seen across different interacting SN types (Figure~\ref{fig:simobj}), with the difference attributed to CSM H/He abundances. For example, SNe~IIn 2009ip and 2019zrk may arise similarly to the scenario considered for SN~2021qqp, where the merger and explosion happen promptly while still embedded in the H-rich CE. With a longer delay to the merger ($\sim10^4$ yr), an unstable Roche-lobe overflow from the WR onto the NS/BH creates a H-poor/He-rich CSM, which may reproduce the precursor and explosion seen in an event like the Type Ibn SN~2022pda. With an even longer delay ($\sim10^5$ yr), the shock interaction between the post-merger disk wind and pre-merger CSM may result in an event such as the Type Icn SN~2021csp. Precursors for SNe~Icn are yet to be seen, but they have the potential to probe this late merger stage.

\section{Conclusions} 
\label{sec:conc}

We have presented optical photometric and spectroscopic observations of the unusual SN~IIn 2021qqp, covering a year-long precursor preceding the explosion and a second distinct peak about a year after the explosion. The precursor is on the luminous ($-15.8\,{\rm mag}<M_{g,r,i}<-14.5\,{\rm mag}$) and long-lasting ($\sim 300$\,days) ends of the distribution for SN~IIn, suggesting extreme mass-loss event(s). The sharp first maximum ($M_g=-19.4$, $M_r=-19.5$\,mag) is characterized by a concave-up curvature, while the second peak ($M_g=-16.7$, $M_r=-17.3$\,mag) is characterized by a concave-down curvature, with possible hints of additional bumps. Throughout the evolution, the spectra are dominated by Balmer lines, with weaker He~{\sc i}, Na~{\sc i}, Ca~{\sc i}, and Fe~{\sc ii} lines appearing around the second peak. By decomposing the multi-component H$\alpha$ and H$\beta$ lines, the CSM and SN--CSM shell velocities are estimated from the core FWHMs and absorption minima, respectively, as $\approx 1300$ and $8500$\,km\,s$^{-1}$ (first peak) and $\approx 2500$ and $5600$\,km\,s$^{-1}$ (second peak).

Motivate by these observations, we have constructed an analytical model to extract the CSM and SN properties from the bolometric light curve and velocity evolution. We infer the presence of two distinct CSM density peaks resulting from episodic mass loss with $\dot{M}\approx10\,M_\odot\,{\rm yr}^{-1}$ about $0.8$\,yr before explosion and $\dot{M}\approx5\,M_\odot\,{\rm yr}^{-1}$ about $2$\,yr before explosion, with a total $M_{\rm CSM}\approx 2-4\,M_\odot$. Moreover, the light-curve precursor could be explained by the most recent mass-loss episode. By imposing a consistent RS luminosity and velocity evolution with the observations, the SN ejecta mass range is constrained to be $M_{\rm SN}\approx 5-30\,M_\odot$ for an explosion energy range of $E_{\rm SN}\approx (3-10)\times10^{51}$\,erg. 

An eruptive massive star (LBV giant eruption, PPISN) or WR-NS/BH merger may be possible progenitor channels for producing such an energetic explosion in a complex CSM environment. Continued monitoring of SN~2021qqp is necessary to further investigate the progenitor channel. If less luminous light-curve peak(s), corresponding to less dense CSM peak(s), were seen quasi-periodically, the stellar merger scenario with eccentric encounters would be favored. The lack of such light-curve peak(s) or periodicity would instead favor the eruptive stellar activity scenario. 

Finally, we note that given the sharp light-curve morphology, events like SN~2021qqp may appear as FBOTs if only observed near peak, for example, at $z\gtrsim 0.13$ for current transient surveys, such as ZTF, with a typical limiting magnitude of $\approx 21$.  We therefore speculate that some FBOTs may have precursor activity and mass-loss episodes similar to those we infer for SN~2021qqp.  Looking forward, the much deeper observations available from the Vera C.~Rubin Observatory's Legacy Survey of Space and Time ($\approx25$\,mag; \citealt{LSST2019ApJ...873..111I}) will reveal SN~2021qqp-like precursors ($\lesssim-14.5$\,mag) to $z\approx0.17$, providing a much larger sample size with complete light-curve coverage (by a factor of $\approx300$ for volume compared to ZTF), including for FBOTs. Such a large sample of precursor-associated transients coupled with our analysis and modeling frameworks presented here would allow us to systematically explore detailed CSM configurations in a self-consistent way and potentially map them to their progenitor systems.

\section{acknowledgments}

We are grateful to Morgan MacLeod, Brian Metzger, Noam Soker, Dillon Brout, and Floor Broekgaarden for useful discussions; Joel Leja for helpful advice on \texttt{Prospector}; Benjamin Weiner for scheduling the MMT Binospec observations; Yuri Beletsky for performing the Magellan LDSS-3 observations; and Jamison Burke for assisting in scheduling LCO observations. 

The Berger Time-Domain research group at Harvard is supported by the NSF and NASA. 
T.M. is supported in part by JSPS Overseas Research Fellowships.
C.R. and V.A.V. are supported by Charles E. Kaufman Foundation New Investigator grant KA2022-129525.
The LCO group is supported by NSF grants AST-1911151 and AST-1911225. 
This publication was made possible through the support of an LSSTC Catalyst Fellowship to K.A.B., funded through grant 62192 from the John Templeton Foundation to LSST Corporation. The opinions expressed in this publication are those of the authors and do not necessarily reflect the views of LSSTC or the John Templeton Foundation.

Observations reported here were obtained at the MMT Observatory, a joint facility of the Smithsonian Institution and the University of Arizona. 
This paper includes data gathered with the 6.5 meter Magellan Telescopes located at the Las Campanas Observatory, Chile.

This work makes use of observations from the Las Cumbres Observatory global telescope network. This paper is based in part on observations made with the MuSCAT3 instrument, developed by the Astrobiology Center and under financial support by JSPS KAKENHI (grant No. JP18H05439) and JST PRESTO (grant No. JPMJPR1775), at Faulkes Telescope North on Maui, HI, operated by the Las Cumbres Observatory. The authors wish to recognize and acknowledge the very significant cultural role and reverence that the summit of Haleakal$\bar{\text{a}}$ has always had within the indigenous Hawaiian community. We are most fortunate to have the opportunity to conduct observations from the mountain. 

This work has made use of data from the Zwicky Transient Facility (ZTF).  ZTF is supported by NSF grant No. AST-1440341 and a collaboration including Caltech, IPAC, the Weizmann Institute for Science, the Oskar Klein Center at Stockholm University, the University of Maryland, the University of Washington, Deutsches Elektronen-Synchrotron and Humboldt University, Los Alamos National Laboratories, the TANGO Consortium of Taiwan, the University of Wisconsin at Milwaukee, and Lawrence Berkeley National Laboratories. Operations are conducted by COO, IPAC, and UW. The ZTF forced-photometry service was funded under the Heising-Simons Foundation grant
No. 12540303 (PI: Graham).

ALeRCE is an initiative funded by the Millennium Institute for Astrophysics – MAS, the Center for Mathematical Modeling - CMM at Universidad de Chile, and since 2020 the Data Observatory, in collaboration with researchers from Universidad Adolfo Ib{\'a}{\~n}ez - UAI, Universidad Austral de Chile - UACH (Informatics), Universidad Catlica de Chile - UC (Astronomy), Universidad de Chile - UCH (Astronomy - DAS, Electrical Engineering - DIE), Universidad de Concepci{\'o}n - UdeC (Informatics), Universidad Nacional Andres Bello – UNAB (Astronomy), Universidad de Santiago de Chile - USACH (Statistics), Universidad Tecnol{\'o}gica Metropolitana - UTEM (Computer Science), Universidad de Valpara{\'i}so - UV (Astronomy), and REUNA in Chile, and international researchers from Caltech and Harvard U. and U. of Washington.

This work has made use of data from the Asteroid Terrestrial-impact Last Alert System (ATLAS) project. ATLAS is primarily funded to search for near-Earth asteroids through NASA grant Nos. NN12AR55G, 80NSSC18K0284, and 80NSSC18K1575; byproducts of the NEO search include images and catalogs from the survey area. This work was partially funded by Kepler/K2 grant No. J1944/80NSSC19K0112 and HST grant No. GO-15889, and STFC grant Nos. ST/T000198/1 and ST/S006109/1. The ATLAS science products have been made possible through the contributions of the University of Hawaii Institute for Astronomy, the Queen’s University Belfast, the Space Telescope Science Institute, the South African Astronomical Observatory, and The Millennium Institute of Astrophysics (MAS), Chile.

The PS1 and the PS1 public science archives have been made possible through contributions by the Institute for Astronomy, the University of Hawaii, the Pan-STARRS Project Office, the Max-Planck Society and its participating institutes, the Max Planck Institute for Astronomy, Heidelberg and the Max Planck Institute for Extraterrestrial Physics, Garching, The Johns Hopkins University, Durham University, the University of Edinburgh, the Queen's University Belfast, the Harvard-Smithsonian Center for Astrophysics, the Las Cumbres Observatory Global Telescope Network Incorporated, the National Central University of Taiwan, the Space Telescope Science Institute, NASA under grant No. NNX08AR22G issued through the Planetary Science Division of the NASA Science Mission Directorate, NSF grant No. AST-1238877, the University of Maryland, Eotvos Lorand University, the Los Alamos National Laboratory, and the Gordon and Betty Moore Foundation.

This publication makes use of data products from the Wide-field Infrared Survey Explorer, which is a joint project of the University of California, Los Angeles, and the Jet Propulsion Laboratory/California Institute of Technology, funded by the National Aeronautics and Space Administration.

This work made use of data supplied by the UK Swift Science Data Centre at the University of Leicester.

The National Radio Astronomy Observatory is a facility of the National Science Foundation operated under cooperative agreement by Associated Universities, Inc.

The Legacy Surveys consist of three individual and complementary projects: the Dark Energy Camera Legacy Survey (DECaLS; Proposal ID \#2014B-0404; PIs: David Schlegel and Arjun Dey), the Beijing-Arizona Sky Survey (BASS; NOAO Prop. ID \#2015A-0801; PIs: Zhou Xu and Xiaohui Fan), and the Mayall z-band Legacy Survey (MzLS; Prop. ID \#2016A-0453; PI: Arjun Dey). DECaLS, BASS and MzLS together include data obtained, respectively, at the Blanco telescope, Cerro Tololo Inter-American Observatory, NSF’s NOIRLab; the Bok telescope, Steward Observatory, University of Arizona; and the Mayall telescope, Kitt Peak National Observatory, NOIRLab. The Legacy Surveys project is honored to be permitted to conduct astronomical research on Iolkam Du’ag (Kitt Peak), a mountain with particular significance to the Tohono O’odham Nation.

NOIRLab is operated by the Association of Universities for Research in Astronomy (AURA) under a cooperative agreement with the National Science Foundation.

This project used data obtained with the Dark Energy Camera (DECam), which was constructed by the Dark Energy Survey (DES) collaboration. Funding for the DES Projects has been provided by the U.S. Department of Energy, the U.S. National Science Foundation, the Ministry of Science and Education of Spain, the Science and Technology Facilities Council of the United Kingdom, the Higher Education Funding Council for England, the National Center for Supercomputing Applications at the University of Illinois at Urbana-Champaign, the Kavli Institute of Cosmological Physics at the University of Chicago, Center for Cosmology and Astro-Particle Physics at the Ohio State University, the Mitchell Institute for Fundamental Physics and Astronomy at Texas A\&M University, Financiadora de Estudos e Projetos, Fundacao Carlos Chagas Filho de Amparo, Financiadora de Estudos e Projetos, Fundacao Carlos Chagas Filho de Amparo a Pesquisa do Estado do Rio de Janeiro, Conselho Nacional de Desenvolvimento Cientifico e Tecnologico and the Ministerio da Ciencia, Tecnologia e Inovacao, the Deutsche Forschungsgemeinschaft and the Collaborating Institutions in the Dark Energy Survey. The Collaborating Institutions are Argonne National Laboratory, the University of California at Santa Cruz, the University of Cambridge, Centro de Investigaciones Energeticas, Medioambientales y Tecnologicas-Madrid, the University of Chicago, University College London, the DES-Brazil Consortium, the University of Edinburgh, the Eidgenossische Technische Hochschule (ETH) Zurich, Fermi National Accelerator Laboratory, the University of Illinois at Urbana-Champaign, the Institut de Ciencies de l’Espai (IEEC/CSIC), the Institut de Fisica d’Altes Energies, Lawrence Berkeley National Laboratory, the Ludwig Maximilians Universitat Munchen and the associated Excellence Cluster Universe, the University of Michigan, NSF’s NOIRLab, the University of Nottingham, the Ohio State University, the University of Pennsylvania, the University of Portsmouth, SLAC National Accelerator Laboratory, Stanford University, the University of Sussex, and Texas A\&M University.

BASS is a key project of the Telescope Access Program (TAP), which has been funded by the National Astronomical Observatories of China, the Chinese Academy of Sciences (the Strategic Priority Research Program “The Emergence of Cosmological Structures” Grant \# XDB09000000), and the Special Fund for Astronomy from the Ministry of Finance. The BASS is also supported by the External Cooperation Program of Chinese Academy of Sciences (Grant \# 114A11KYSB20160057), and Chinese National Natural Science Foundation (Grant \# 11433005).

The Legacy Survey team makes use of data products from the Near-Earth Object Wide-field Infrared Survey Explorer (NEOWISE), which is a project of the Jet Propulsion Laboratory/California Institute of Technology. NEOWISE is funded by the National Aeronautics and Space Administration.

The Legacy Surveys imaging of the DESI footprint is supported by the Director, Office of Science, Office of High Energy Physics of the U.S. Department of Energy under Contract No. DE-AC02-05CH1123, by the National Energy Research Scientific Computing Center, a DOE Office of Science User Facility under the same contract; and by the U.S. National Science Foundation, Division of Astronomical Sciences under Contract No. AST-0950945 to NOAO.

This research has made use of the NASA Astrophysics Data System (ADS), the NASA/IPAC Extragalactic Database (NED), and NASA/IPAC Infrared Science Archive (IRSA, which is funded by NASA and operated by the California Institute of Technology), and IRAF (which is distributed by the National Optical Astronomy Observatory, NOAO, operated by the Association of Universities for Research in Astronomy, AURA, Inc., under cooperative agreement with the NSF).

TNS is supported by funding from the Weizmann Institute of Science, as well as grants from the Israeli Institute for Advanced Studies and the European Union via ERC grant No. 725161.

We acknowledge with thanks the variable star observations from the AAVSO International Database contributed by observers worldwide and used in this research.


\vspace{5mm}
\facilities{AAVSO, ADS, ATLAS, DECam, GALEX, IRSA, Keck (LRIS), LCO (MuSCAT3, Sinistro), Legacy Surveys, Magellan (LDSS-3), MMT (Binospec), NTT (EFOSC2), NED, PS1, PTF, SDSS, \textit{Swift} (XRT, UVOT), VLASS, WISE, ZTF}.

\defcitealias{Astropy2018AJ....156..123A}{Astropy Collaboration 2018}
\software{Astropy \citepalias{Astropy2018AJ....156..123A}, 
\texttt{atlas-fp} (\url{https://gist.github.com/thespacedoctor/86777fa5a9567b7939e8d84fd8cf6a76}), 
BANZAI \citep{curtis_mccully_2018_1257560},
\texttt{dynesty} \citep{Speagle2020MNRAS.493.3132S},
\texttt{emcee} \citep{Foreman-Mackey2013PASP..125..306F},
\texttt{FSPS} \citep{Conroy2009ApJ...699..486C,Conroy2010ApJ...712..833C},
\texttt{lcogtsnpipe} \citep{Valenti2016MNRAS.459.3939V}, 
Matplotlib \citep{Hunter2007CSE.....9...90H}, 
NumPy \citep{Oliphant2006}, 
\texttt{photutils} \citep{photutils},
PyRAF \citep{PyRAF2012ascl.soft07011S},
\texttt{Prospector} \citep{Johnson2021ApJS..254...22J},
\texttt{pwkit} \citep{Williams2017ApJ...834..117W},
SciPy \citep{SciPy2020NatMe..17..261V},
seaborn \citep{seaborn2020zndo...3629446W},
\texttt{sedpy} \citep{sedpy},
\texttt{SExtractor} \citep{Bertin1996A&AS..117..393B}}.

\bibliography{main,reference_matsumoto}

\clearpage


\end{document}